\documentclass[aps,prl,twocolumn,groupedaddress,nofootinbib]{revtex4-1}
\usepackage{amssymb}
\usepackage[a4paper, total={6in, 8.3in}]{geometry}
\usepackage{sidecap}
\usepackage{float}
\usepackage{multirow}
\usepackage[utf8]{inputenc}
\usepackage{latexsym}
\usepackage{enumerate}
\usepackage{caption}
\usepackage[hyperref]{xcolor}
\usepackage{epsfig}
\graphicspath{{projectplots/}}
\renewcommand\thefootnote{\textcolor{red}{\fnsymbol{footnote}}}

\usepackage{relsize}
\usepackage{indentfirst}
\usepackage{amssymb}
\usepackage{bookmark}
\usepackage{amsthm}
\usepackage{amsmath}

\usepackage{tabularx} 
\usepackage{amsmath}  
\usepackage{graphicx} 
\usepackage{hyperref} 

\definecolor{nicered}{rgb}{0.7,0.1,0.1}
\definecolor{MyDarkBlue}{rgb}{0,0.1,0.7}
\definecolor{DarkBlue}{RGB}{0,0,153}
\definecolor{PetuniaColor}{RGB}{150,23,147}
\definecolor{Purple}{RGB}{128,0,128}

\hypersetup{colorlinks,bookmarksopen,bookmarksnumbered,
citecolor={nicered},
linkcolor={MyDarkBlue},pdfstartview=FitH,
urlcolor={DarkBlue}}

\usepackage{titlesec}
\titlespacing*{\section}
{0pt}{1\baselineskip}{0.8\baselineskip}

\usepackage{scalerel}
\usepackage{tikz}
\usetikzlibrary{svg.path}
\definecolor{orcidlogocol}{HTML}{A6CE39}
\tikzset{
  orcidlogo/.pic={
    \fill[orcidlogocol] svg{M256,128c0,70.7-57.3,128-128,128C57.3,256,0,198.7,0,128C0,57.3,57.3,0,128,0C198.7,0,256,57.3,256,128z};
    \fill[white] svg{M86.3,186.2H70.9V79.1h15.4v48.4V186.2z}
                 svg{M108.9,79.1h41.6c39.6,0,57,28.3,57,53.6c0,27.5-21.5,53.6-56.8,53.6h-41.8V79.1z M124.3,172.4h24.5c34.9,0,42.9-26.5,42.9-39.7c0-21.5-13.7-39.7-43.7-39.7h-23.7V172.4z}
                 svg{M88.7,56.8c0,5.5-4.5,10.1-10.1,10.1c-5.6,0-10.1-4.6-10.1-10.1c0-5.6,4.5-10.1,10.1-10.1C84.2,46.7,88.7,51.3,88.7,56.8z};}}
\newcommand\orcid[1]{\href{https://orcid.org/#1}{\mbox{\scalerel*{
\begin{tikzpicture}[yscale=-1,transform shape]
\pic{orcidlogo};
\end{tikzpicture}
}{|}}}}

\def\beq{\begin{equation}}
\def\eeq{\end{equation}}
\def\bea{\begin{eqnarray}}
\def\eea{\end{eqnarray}}

\newcommand{\newc}{\newcommand}
\def\eq$#1${\begin{equation}#1\end{equation}}
\def\gat$#1${\begin{gather}#1\end{gather}}
\def\bal$#1${\begin{align}#1\end{align}}
\def\eqarr$#1${\begin{eqnarray}#1\end{eqnarray}}
\newc{\pa}{\partial}
\newc{\alp}{\alpha}
\newc{\gam}{\gamma}
\newc{\Gam}{\Gamma}
\newc{\del}{\delta}
\newc{\eps}{\epsilon}
\newc{\lam}{\lambda}
\newc{\sig}{\sigma}
\newc{\ups}{\upsilon}
\newc{\ome}{\omega}
\newc{\pphi}{\varphi}
\newc{\nonum}{\nonumber}
\newc{\hami}{\text{\textbf{\lat{H}}}}
\newc{\gren}{\mathcal{G}}
\newc{\lagr}{\mathcal{L}}
\newc{\timor}{\mathcal{T}}
\newc{\prop}{\mathcal{K}}
\newc{\zcal}{\mathcal{Z}}
\newc{\cinf}{\mathcal{C}_\infty}
\newc{\operx}{\text{\textbf{\lat{x}}}}
\newc{\opera}{\text{\textbf{\lat{a}}}}
\newc{\operp}{\text{\textbf{\lat{p}}}}
\newc{\operl}{\text{\textbf{\lat{L}}}}
\newc{\gfv}{g^{(5)}}
\newc{\kfv}{\kappa_{(5)}}
\newc{\tf}{\tilde{f}}
\newc{\tlam}{\tilde{\Lambda}}
\newc{\tl}{\tilde{\lam}}
\newc{\dist}{\displaystyle}
\newc{\ra}{\rightarrow}
\newc{\Ra}{\Rightarrow}
\newc{\hsp}{\hspace{1em}}

\begin{document}


\title{Localized brane-world black hole analytically connected
to an AdS$_5$ boundary}

\author{\textbf{Theodoros Nakas}\orcid{0000-0002-3522-5803}}
\email[]{\,t.nakas@uoi.gr}
\author{\textbf{Panagiota Kanti}\orcid{0000-0002-3018-5558}}
\email[]{\,pkanti@uoi.gr}
\affiliation{Division of Theoretical Physics, Department of Physics,
             University of Ioannina, GR-45110, Greece}

\begin{abstract}
We construct from first principles the geometry of an analytic, exponentially localized five-dimensional
brane-world black hole. The black-hole singularity lies entirely on the 3-brane, while the event horizon
is shown to have a pancake shape. The induced line-element on the brane assumes the form of the
Schwarzschild solution while the bulk geometry is effectively AdS$_5$ outside the horizon.
The derived geometry is supported by an anisotropic fluid in the bulk described only by two independent 
components, the energy density and tangential pressure, whereas no matter needs to be introduced on
the brane for its consistent embedding in the bulk.
\end{abstract}
\maketitle


\renewcommand*{\thefootnote}{\arabic{footnote}}

\section{INTRODUCTION}

The warped brane-world model \cite{RS1, RS2} made its appearance more than twenty years ago 
and radically changed the way we thought about theories with extra spatial dimensions. In this,
our four-dimensional world is a flat 3-brane \cite{misha, akama} embedded in a five-dimensional Anti-de Sitter (AdS)
spacetime supported by a negative bulk cosmological constant. The length of the fifth extra
dimension can be either finite or infinite since the localisation of the graviton close
to our brane is ensured by an exponentially decaying warp factor multiplying the
line-element of the four-dimensional brane manifold. 

The question of whether the flat background on the brane could be replaced by a curved one
soon emerged. In \cite{CHR}, the four-dimensional Minkowski line-element was replaced by the
Schwarzschild one in an attempt to construct a black hole localised on the brane, as a brane
observer would witness. However, this led instead to an infinitely-long, unstable \cite{GL, Ruth}
black-string solution plagued also by a curvature singularity at the AdS infinity. Adopting
a different perspective where the bulk geometry played the leading role, numerical solutions
describing small \cite{Kudoh1, Kudoh2} and large \cite{Tanahashi,Figueras, Page} black holes
were constructed. Despite the countless attempts in the literature \cite{EHM, Dadhich, tidal, Papanto, 
Bruni, KT, Dadhich2, CasadioNew, EFK, Frolov, EGK, Tanaka, KOT,Charmousis, Kofinas, Shanka, Karasik, 
GGI, CGKM, Ovalle, Harko, daRocha1, AS, Fitzpatrick, Zegers, Heydari, Dai,   Yoshino, Kleihaus,
daRocha2, Andrianov1, Andrianov2, Cuadros, KPZ, KPP}, adopting either the
brane or the bulk perspective, no analytic solution has been constructed so far describing
a five-dimensional black-hole line-element localised close to our brane and leading also to a
Schwarzschild black hole on the brane. 

This is the task we have undertaken in the present work. We are adopting a combined bulk/brane
perspective in which we construct from first principles the geometry of an analytic, spherically-symmetric
five-dimensional black hole which also reduces to a Schwarzschild black hole on the brane. We
demonstrate that the singularity remains strictly localised on the brane thus avoiding the
emergence of any singularities in the bulk. The horizon has a pancake shape as it is localised 
exponentially close to our brane, while the five-dimensional background
quickly reduces to a pure AdS$_5$ spacetime away from the brane. In order to
support this well-behaved geometry, we need an anisotropic fluid in the bulk described only
by two independent components, the energy density and tangential pressure, which satisfy all
energy conditions on the brane. We finally demonstrate that no matter needs to be introduced on
the brane for its consistent embedding in the bulk and that the effective theory on the brane
is satisfied by the four-dimensional Schwarzschild geometry as expected.

\section{THE GEOMETRY}

The line-element of the Randall-Sundrum model \cite{RS1, RS2} has the following well-known form 
\eq$\label{rs-metric1}
ds^2=e^{-2k|y|}\left(-dt^2+d\vec x ^{\,2}\right)+dy^2\,.$
It describes a five-dimensional spacetime comprised by four-dimensional flat slices
stacked together along a fifth dimension denoted by the coordinate $y$. Each slice has
a warp factor $e^{-2k|y|}$, where $k$ is the curvature of the five-dimensional
Anti-de Sitter (AdS) spacetime supported by a negative bulk cosmological constant.  
A 3-brane must be introduced at the location $y=0$: the cusp in the first derivative of
the warp factor, and thus of the metric tensor, demands the presence of some distribution
of matter at this point, which makes this slice of the AdS spacetime a physical boundary---our 
four-dimensional world. 

The line-element \eqref{rs-metric1} may be alternatively written in conformally-flat coordinates:
introducing the new coordinate $z$ via the relation $z=sgn(y)\,(e^{k |y|}-1)/k$, this takes the form
\eq$\label{rs-metr-z}
ds^2=\frac{1}{(k|z|+1)^2}\left(-dt^2+dr^2+r^2\,d\Omega_2^2+dz^2\right)\,.$
In the above expression, we have also employed spherical coordinates for the spatial directions
on the brane with $d\Omega_2^2=d\theta^2+ \sin^2\theta\,d\phi^2$. We note that, in terms of the new
bulk coordinate, the location of the brane is also at
$z=0$; there, the value of the warp factor is equal to unity. At the AdS asymptotic boundary, i.e. at
$|y| \rightarrow \infty$ or $|z| \rightarrow \infty$, the warp factor vanishes. In the case of
the regular spacetime \eqref{rs-metric1} this signifies merely a coordinate singularity,
however, when combined with a black-hole line-element on the brane, it leads to a true
spacetime singularity at the AdS infinity \cite{CHR}. 

\begin{figure}[t!]
\hspace*{-2em}
\includegraphics[width=0.55 \textwidth]{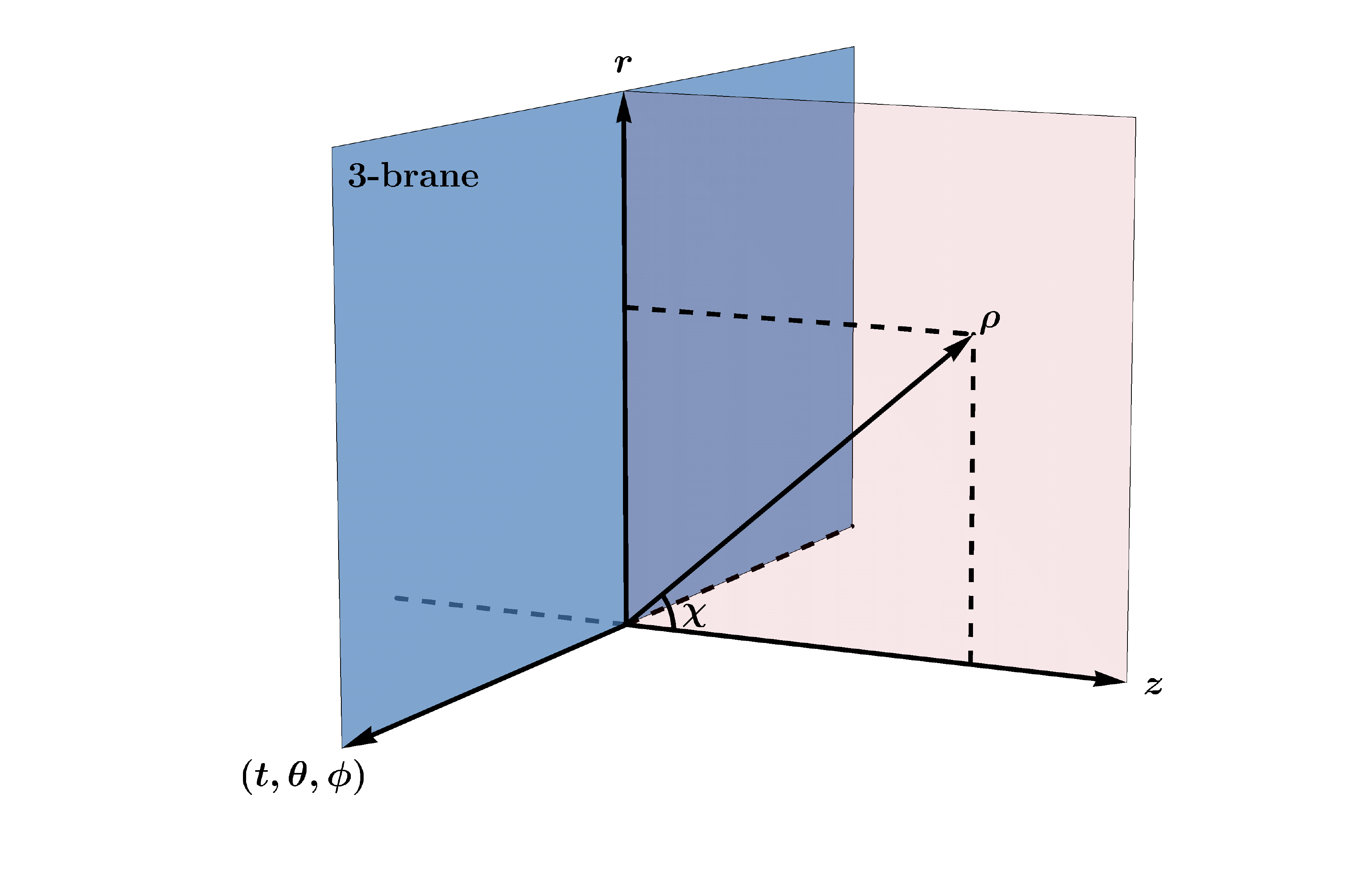}
\vspace{-2.5em} 
 \caption{The geometrical set-up of the five-dimensional spacetime and the set of coordinates.}
\vspace{-0.5em}   
    \label{coords}
\end{figure}

We will now introduce five-dimensional spherical symmetry. To this end, we perform the
following change of variables:
\eq$\label{new-coords}
\left\{\begin{array}{l} r=\rho\,\sin\chi\\[2mm]
z=\rho\,\cos\chi\end{array}\right\}\,, \hspace{1em}{\rm where} \hspace{1em}
\chi\in\left[0,\pi\right]\,.$
Employing these in Eq. \eqref{rs-metr-z}, we obtain
\eq$\label{rs-metr-sph}
ds^2=\frac{1}{(1+k\rho|\cos\chi|)^2}\left(-dt^2+d\rho^2+\rho^2\,d\Omega_3^2\right)\,,$
where $d\Omega_3^2$ is now the line-element of a unit three-dimensional sphere, namely
\eq$
d\Omega_3^2=d\chi^2+\sin^2\chi\,d\theta^2+\sin^2\chi\,\sin^2\theta\,d\phi^2\,.$
The inverse transformation reads
\eq$\label{new-coords-inv}
\Bigl\{\rho= \sqrt{r^2+z^2}\,, \quad
\tan \chi=r/z\Bigr\}\,.$
As Eq. (\ref{new-coords}) dictates, the new radial coordinate $\rho$ is always positive definite
since $\sin\chi \geq 0$ for $\chi\in\left[0,\pi\right]$. On the other hand, $\cos\chi>0$ for $\chi\in\left[0,\pi/2\right)$
and $\cos\chi<0$ for $\chi\in\left(\pi/2,\pi\right]$; thus, the first regime corresponds to positive $z$ and 
describes the bulk spacetime on the right-hand-side of the 3-brane, while the second regime corresponds
to negative $z$ and describes the bulk spacetime on the left-hand-side of the brane.  However, the
corresponding line-elements are related by the coordinate transformation $\chi \rightarrow \pi - \chi$ and thus
describe the same spacetime. The brane itself is located at $\cos \chi=0$, i.e. at $\chi =\pi/2$. 
Fig. \ref{coords} depicts the  geometrical setup of the five-dimensional spacetime. 

The radial coordinate $\rho$ ranges over the interval $[0, \infty)$ since, according to Eq. (\ref{new-coords-inv}),
receives contributions both from the (brane) $r$ and (bulk) $z$ coordinates. Therefore, the
5-dimensional radial infinity, $\rho \rightarrow \infty$, may describe both the asymptotic AdS boundary
($|z| \rightarrow \infty$) and the radial infinity on the brane ($r \rightarrow \infty$). 
We note that, on the brane where $z=0$, $\rho$ reduces to the brane radial coordinate $r$. Due to
the aforementioned symmetry of the line-element under $\chi \rightarrow \pi - \chi$, it is adequate
to consider only one of the two $\chi$-regimes; thus, we henceforth focus on the regime
$\chi\in\left[0,\pi/2\right]$ for which $\cos\chi \geq 0$. 

Inspired by \cite{CHR}, we now replace the two-dimensional flat part $(-dt^2 +d\rho^2)$
of the line-element in Eq. \eqref{rs-metr-sph} with the corresponding part of
the four-dimensional Schwarzschild solution. Thus, we obtain the following five-dimensional
spacetime 
\eq$\label{5d-schw-metr}
ds^2=\frac{1}{(1+k\rho \cos\chi)^2}\bigg[-f(\rho)\,dt^2+\frac{d\rho^2}{f(\rho)}+
\rho^2\,d\Omega_3^2\bigg],$
where $f(\rho)=1-\frac{2M}{\rho}$. At the location of the brane ($\chi=\pi/2$), $\cos \chi=0$ 
and the warp factor reduces to unity. 
Since there it also holds that $\rho=r$, the line-element on the brane reduces to the usual
four-dimensional Schwarzschild solution with the horizon located at $r=r_h=2M$. But what kind
of five-dimensional gravitational background does the line-element \eqref{5d-schw-metr} describe?
To answer this question, we need to evaluate the five-dimensional curvature invariant quantities.
For instance, the five-dimensional Ricci scalar is found to have the form
\eq$\label{ricci-1}
R=-20 k^2 + \frac{12 k^2 M \cos^2 \chi}{\rho }-\frac{24 k M \cos \chi }{\rho ^2}
+\frac{4 M}{\rho ^3}\,.$
The above expression contains a constant contribution $-20 k^2$, attributed to the negative
cosmological constant in the bulk, plus additional terms sourced by the mass $M$ located on
the brane. These terms are inversely proportional to powers of the bulk radial coordinate $\rho$
and  therefore diverge when $\rho \rightarrow 0$. However, since $\rho = \sqrt{r^2 + z^2}$, this 
is realised only when $z=0$ (i.e. on the brane) {\it and} $r=0$ (i.e. at the location of the 
mass $M)$. At any other point in the bulk, characterised by definition by a {\it non-vanishing} value
of $z$, the limit $r \rightarrow 0$ does not lead to a singularity. As a result, the singularity at
$\rho=0$ remains restricted on the brane at the single point $r=0$, and the
line-element \eqref{5d-schw-metr} describes a regular spacetime, which
is analytically connected to a black-hole spacetime on the 3-brane.

In addition, an asymptotically AdS$_5$ spacetime readily emerges when $\rho \rightarrow \infty$:
when we move either far away from the brane, i.e. $z \rightarrow \infty$, or at large distances
along the brane, i.e. $r \rightarrow \infty$, all $\rho$-dependent terms in Eq. (\ref{ricci-1})
vanish. A similar behaviour is found for the other two five-dimensional curvature invariants 
$R_{MN} R^{MN}$ and $R_{MNKL} R^{MNKL}$, the  explicit expressions of which can be found
in Appendix A. Taking  the limit $\rho \rightarrow \infty$, we find the following
asymptotic values for the three invariant quantities:
\begin{eqnarray}
&R=-20 k^2\,,&\\[2mm]
&R_{MN}\,R^{MN}=80 k^4\,,&\\[2mm]
&R_{MNKL}\,R^{MNKL}=40 k^4\,,&
\end{eqnarray}
which match the ones for a pure five-dimensional AdS spacetime. Therefore, the line-element
\eqref{5d-schw-metr} describes in particular a regular asymptotically AdS$_5$ spacetime, which
is analytically connected to a black-hole spacetime on the 3-brane.

\begin{figure}[t!]
    \centering
\includegraphics[width=0.45 \textwidth]{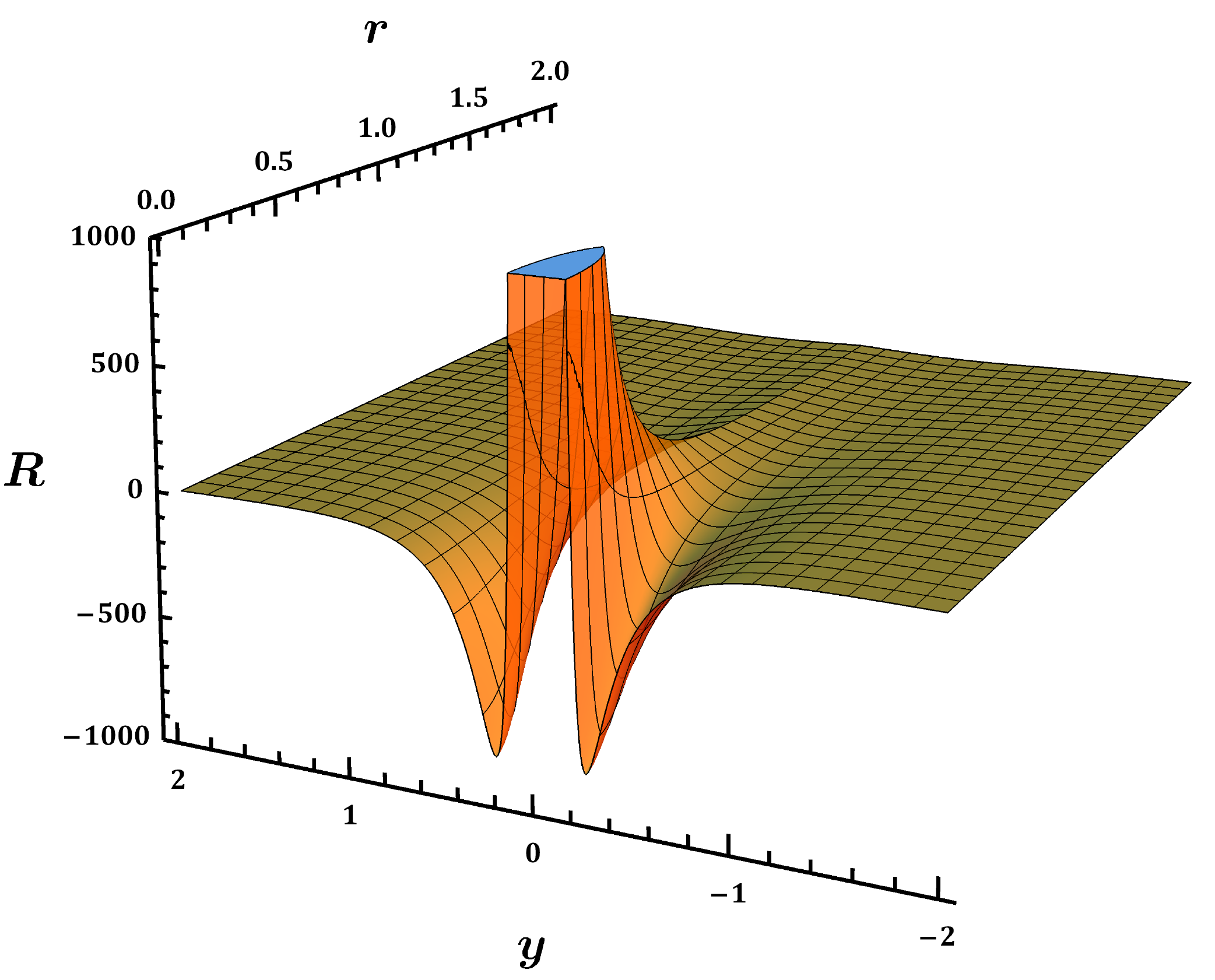}
 \caption{The scalar curvature $R$ in terms of the coordinates $\{r,y\}$ for 
    $k=1$ and $M=10$.}\vspace*{-0.8em}
    \label{Rscalar}
\end{figure}

The complete regularity of the five-dimensional spacetime and the localisation of the black-hole
singularity on the brane are clearly depicted in the profile of the scalar curvature $R$ presented in
Fig. \ref{Rscalar}. In order to obtain a better understanding of the spacetime
geometry, we have retorted to the original $(r,y)$ coordinates making use of Eq. \eqref{new-coords-inv} 
and the relation $z=sgn(y)\,(e^{k |y|}-1)/k$. In terms of these, the Ricci scalar $R$ reads
\eq$\label{ricci1-ry}
R=-20k^2+\frac{4k^3M\,\left(10- 12\,e^{k \left| y\right| }+3\,e^{2 k \left| y\right| }\right)}
{\left[\left(e^{k \left| y\right| }-1\right)^2+ k^2 r^2\right]^{3/2}}\,.$
In Fig. \ref{Rscalar}, we present the overall behaviour of the Ricci scalar
in terms of both $(r,y)$-coordinates; clearly, the singularity arises at $r=0$ if and only if 
$y=0$, too, i.e. at the location of the brane. At every other point of either the bulk or the
brane, the spacetime remains regular. We should note here that the singularity at the AdS
horizon, i.e. as $|y| \rightarrow \infty$, that plagued the black-string solution of \cite{CHR},
is absent here since the second term in Eq. \eqref{ricci1-ry} vanishes altogether in that
limit. The profiles of the other two invariants, namely $R_{MN} R^{MN}$ and
$R_{MNKL} R^{MNKL}$, are similar to that of $R$ with the only difference being their
monotonic rise close to the singularity on the brane and the absence of the double well
observed in Fig. \ref{Rscalar}.

Let us also re-write the five-dimensional line-element \eqref{5d-schw-metr} in terms
of the original non-spherical coordinates $\{r,y\}$. Employing the inverse transformations
(\ref{new-coords-inv}), the line-element takes the form
\bal$\label{metr-r-y}
ds^2&=e^{-2k|y|}\left\{-f(r,y) dt^2+\frac{dr^2}{r^2+z^2(y)}\biggl[\frac{r^2}{f(r,y)}+z^2(y)\biggr]\right.\nonum\\[1mm]
&\hsp+r^2d\Omega_2^2+\left.\frac{2r z(y)\,e^{k|y|}}{r^2+z^2(y)}\biggl[\frac{1}{f(r,y)}-1\biggr]drdy \right\}\nonum\\[1mm]
&\hsp+\frac{dy^2}{r^2+z^2(y)}\left[r^2+\frac{z^2(y)}{f(r,y)}\right]\,,$
where $z=sgn(y)\,(e^{k |y|}-1)/k$ and 
\eq$\label{f-ry}
f(r,y)=1-\frac{2M}{\sqrt{r^2+z^2(y)}}\,.$
We observe that the aforementioned line-element differs significantly from the factorised
line-element employed in \cite{CHR}, or from non-factorised ones which appeared in a number
of subsequent works \cite{KT, KOT, KPZ, KPP}. It is the gradual construction---via the employment
of the spherically symmetric coordinates of the line-element (\ref{5d-schw-metr})---that has
resulted in the expression \eqref{metr-r-y}.

Although the singularity of the black hole remains localised on the brane as demonstrated
above, the horizon of the black hole does not need to do so; in fact, we expect it to extend
into the bulk. Let us therefore study the causal structure of the bulk spacetime as this is defined
by the light cone. We therefore consider radial null trajectories in the five-dimensional background
\eqref{metr-r-y}. For a fixed value $y=y_0$ of the fifth coordinate, the condition $ds^2=0$, with
 $\theta$ and $\phi$ kept constant,  leads to the result
\eq$ 
\frac{dt}{dr}=\pm\frac{1}{f(r,y_0)}\left[\frac{r^2k^2+f(r,y_0)\left(e^{k|y_0|}-1\right)^2}{r^2k^2+
\left(e^{k|y_0|}-1\right)^2}\right]^{1/2},$
where
\eq$\label{horizon-function}
f(r,y_0)=1-2M\left[r^2+\frac{\left(e^{k|y_0|}-1\right)^2}{k^2}\right]^{-1/2}.$
At large distances along the brane, i.e. when $r\ra \infty$, one gets 
$\lim_{r\ra \infty}f(r,y_0)=1$, and the slope ${dt}/{dr}$ goes to $\pm 1$,
as expected. However, at a finite distance $r=r_h$, which is defined through the relation
$f(r_h,y_0)=0$, we obtain $dt/dr=\pm\infty$. Therefore, at $y=y_0$ and $r=r_h$ we
encounter the horizon of the black hole as it extends into the bulk. Its exact location follows
from Eq. \eqref{horizon-function},  and is given by
\eq$\label{r-hor}
r_h^2=4M^2-\frac{\left(e^{k|y_0|}-1\right)^2}{k^2}\,.$
We note that $r_h$ depends on the parameters $M$, $k$ and $y_0$.
On the 3-brane, where $y_0=0$, $r_h$ equals the Schwarzschild value $2M$ independently
of the value of $k$, in agreement with the discussion above Eq. \eqref{ricci-1}. 
But, as we move along the extra dimension, $r_h$ shrinks exponentially fast and
becomes zero at a distance 
\eq$|y_0|=\frac{1}{k}\ln\left(2Mk+1\right)$ 
away from the brane. As a result, the black-hole horizon has the shape
of a ``pancake'' with its long side lying along the brane and its short side extending in the
bulk over an exponentially small distance.  In conclusion, the line-element  \eqref{metr-r-y}---or 
its spherically-symmetric analog (\ref{5d-schw-metr})---describes a five-dimensional
black-hole solution which exhibits a localization of its singularity {\it  strictly on our brane}
and a localisation of its horizon  {\it exponentially close} to our brane.


\section{THE GRAVITATIONAL THEORY}

We now turn to the gravitational theory and consider the following five-dimensional action functional
\eq$\label{bulk-action}
S_{B}=\int d^5x\, \sqrt{-g} \left(\frac{R}{2\kappa_5^2}+\lagr^{(B)}_{m}\right).$
In the above, $g_{MN}$ is the metric tensor of the five-dimensional spacetime, $\kappa_5^2=8\pi G_5$ 
incorporates the five-dimensional gravitational constant $G_5$, while $R$ is the five-dimensional Ricci
scalar. The Lagrangian  density $\lagr^{(B)}_m$ describes the matter that exists in the bulk. 

\par The gravitational field equations in the bulk can be obtained by the variation of the action  $S_B$
with respect to the metric tensor $g_{MN}$. These read
\eq$\label{field-eqs}
G_{MN}=\kappa_5^2\, T^{(B)}_{MN}= -\frac{2 \kappa_5^2}{\sqrt{-g}}\frac{\delta\left(\lagr^{(B)}_{m}\sqrt{-g}\right)}{\delta g^{MN}}\,,$
where $G_{MN}=R_{MN}-\frac{1}{2}\,g_{MN}R$ \, is the Einstein-tensor and $T^{(B)}_{MN}$ the 
energy-momentum tensor associated with the Lagrangian density $\lagr^{(B)}_m$. 

\par We will now employ the form of the  gravitational background described by the line-element
(\ref{5d-schw-metr}). Substituting on the left-hand-side of Eq. (\ref{field-eqs}) and solving for the
components of the energy-momentum tensor, we find:
\gat$\label{rho-p1}
T^{(B)t}{}_{t}=T^{(B)\rho}{}_{\rho}=\frac{1}{\kappa_5^2}\biggl(6 k^2+\frac{9 k M \cos \chi}{\rho ^2}
-\frac{3 M}{\rho ^3}\biggr),\\[1mm]
\hspace*{-3.5cm}T^{(B)\chi}{}_{\chi}=T^{(B)\theta}{}_{\theta}=T^{(B)\phi}{}_{\phi} \nonum \\[1mm]
\hspace*{1.3cm}=\frac{1}{\kappa_5^2}\left(6 k^2-\frac{6 k^2 M \cos ^2 \chi }{\rho }+\frac{6 k M \cos \chi}{\rho ^2}\right).
\label{p2}$
Therefore, the gravitational background (\ref{5d-schw-metr}) may be supported by a bulk energy-momentum tensor
the only non-vanishing components of which are the energy-density $\rho_E \equiv -T^{(B)t}{}_{t}$, the radial pressure
$p_1 \equiv T^{(B)\rho}{}_{\rho}$,  and a common tangential pressure $p_2 \equiv T^{(B)\chi}{}_{\chi}=T^{(B)\theta}{}_{\theta}=
T^{(B)\phi}{}_{\phi}$. The necessary matter content of the bulk is thus an anisotropic fluid described by a {\it diagonal}
energy-momentum tensor which in a covariant notation may be written as 
\eq$\label{en-mom}
T^{(B)MN}=(\rho_E+p_2)U^M U^N+(p_1-p_2)X^M X^N+p_2\, g^{MN}\,.$
In the above expression,  $U^M$ is the fluid's timelike five-velocity, and $X^M$ is a 
spacelike unit vector in the direction of $\rho$-coordinate satisfying the relations
\gat$
U^M=\{U^t,0,0,0,0\},\hspace{1em} U^M U^N g_{MN}=-1\,,\\[2mm]
X^M=\{0,X^\rho,0,0,0\},\hspace{1em} X^M X^N g_{MN}=1\,.$

We observe that the energy density and radial pressure satisfy the equation of state $p_1=w_1\, \rho_E$ with $w_1=-1$
everywhere in the bulk, whereas $w_2$ defined via $p_2=w_2\, \rho_E$, is $(\rho,\chi)$-dependent. However, we
note that as $\rho\ra+\infty$,  all components of the energy-momentum
tensor reduce to a constant which can be identified as the five-dimensional cosmological constant $\Lambda_5$
\gat$
\lim_{\rho\ra+\infty}\rho_{E}(\rho,\chi)=-\frac{6k^2}{\kappa_5^2} \equiv \Lambda_5\,, \\[1mm]
\lim_{\rho\ra+\infty}p_1(\rho,\chi)=\lim_{\rho\ra+\infty}p_2(\rho,\chi)=\frac{6k^2}{\kappa_5^2} \equiv
-\Lambda_5\,.$
This negative cosmological constant supports the AdS$_5$ gravitational background away from the brane,
in agreement with the derived asymptotic behaviour of the line-element (\ref{5d-schw-metr}). It also leads to
the exponentially decaying warp factor $e^{-k |y|}$ of the bulk spacetime and thus incorporates the
Randall-Sundrum model \cite{RS1, RS2}. 

\begin{figure}
\begin{center}
    \includegraphics[width=0.48\textwidth]{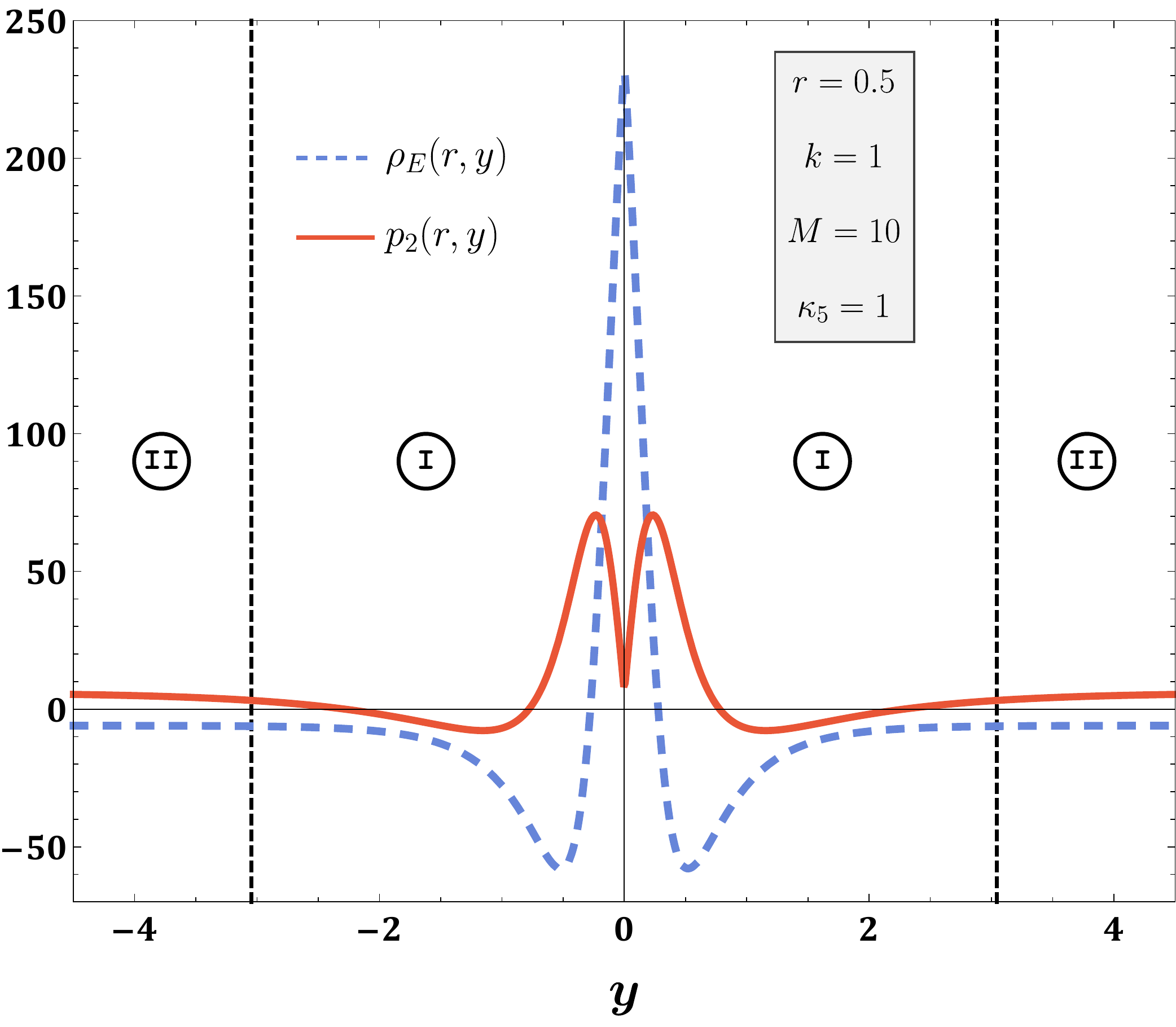}
    \end{center}
    \vspace{-1.5em}
    \caption{The $\rho_E(r,y)$ and $p_2(r,y)$ in terms of the $y$-coordinate for $r=0.5$, $k=1$, $M=10$ and $\kappa_5=1$. Region I
     lies inside the black-hole horizon, while region II lies outside the black-hole horizon. \vspace{-0.5em}}
    \label{rho-p2}  
\end{figure}


\par Let us now study the profiles of the energy density $\rho_E$ and tangential pressure $p_2$ of the bulk
anisotropic fluid. To this end, we employ the coordinates $\{r, y\}$ in terms of which their characteristics are
more transparent.  Using Eqs. (\ref{new-coords-inv}), (\ref{rho-p1}) and (\ref{p2}), we find
\gat$\label{rho-p1-ry}
\rho_E=\frac{3k^2}{\kappa_5^2}\Biggl\{-2+\frac{kM \left(4-3\, e^{k|y|}\right)}{\left[k^2r^2+\left(e^{k|y|}-1\right)^2
\right]^{3/2}}\Biggr\},\\[1mm]
p_2=\frac{6k^2}{\kappa_5^2}\Biggl\{1+\frac{kM\left(e^{k|y|}-1\right)\left(2-e^{k|y|}\right)}
{\left[k^2r^2+\left(e^{k|y|}-1\right)^2\right]^{3/2}}\Biggr\}.$
In Fig. \ref{rho-p2}, we depict the above quantities in terms of the extra dimension $y$, for $r=0.5$,  $k=1$,
$M=10$ and $\kappa_5=1$. We have divided the bulk spacetime in regions I and II; the former lies inside
the black-hole horizon---which for the selected values is located at the bulk coordinate $|y|\simeq 3$---while 
the latter lies outside of it.  It is worth noticing that outside the black-hole horizon both $\rho_E$ and
$p_2$ quickly approach their asymptotic values, which means that the spacetime outside the horizon is
effectively AdS$_5$. On the other hand, on the brane, located at $y=0$, the energy density and tangential
pressure adopt finite values which satisfy all energy conditions since $\rho_E>0$, $\rho_E+p_1=0$
and $\rho_E \gg p_2$. It is only off the brane and within the black-hole horizon that the bulk fluid exhibits
a non-conventional behaviour as revealed by the violation of the energy conditions. This is a necessary
feature for the localisation of the black hole  near the brane, which would otherwise `leak' into the bulk
resulting in a black string \cite{KNP1,KNP2,KNP3}. In fact, one could compare this requirement with the violation 
of the energy conditions around the throat of a wormhole \cite{MT}---there, as well as in our case, the
violation is only  {\it local} and {\it necessary} for the support of the desired geometry.

\section{JUNCTION CONDITIONS AND EFFECTIVE THEORY ON THE BRANE}

Our four-dimensional world is described by the 3-brane ($\Sigma,\, h_{MN}$) embedded in the five-dimensional
spacetime ($\mathcal{M},\, g_{MN}$) at $y=0$.  From Eq. \eqref{metr-r-y} the unit vector normal to the
3-brane is determined to be $n^M=\del^{M}{}_y$, hence, the induced metric on the brane is defined via
$h_{MN}=(g_{MN})|_{y= 0}-n_M\, n_N$.

As in all brane-world scenarios, we decompose the total
energy-momentum tensor as
\eq$\label{ene-mom-dec}
T_{MN}=T^{(B)}_{MN}+\del^{\mu}_M\del^{\nu}_N\,T^{(br)}_{\mu\nu}\del(y)\,,$
where $T^{(B)}_{MN}$ and $T^{(br)}_{\mu\nu}$ denote the bulk and brane energy-momentum tensors,
respectively. The brane  energy-momentum tensor can be broken down further, namely
\eq$\label{brane-ene-dec}
T^{(br)}_{\mu\nu}=-\sigma\, h_{\mu\nu}+\tau_{\mu\nu}\,,$
where $\sigma$ is the vacuum energy (or tension) on the brane, and $\tau_{\mu\nu}$ encodes all the
other possible sources of energy and/or pressure on the brane. In what follows, we will first investigate
whether the consistent embedding of our brane in the five-dimensional line-element (\ref{metr-r-y})
demands the introduction of a non-trivial $\tau_{\mu\nu}$. 

To this end, we will study Israel's junction conditions at $y=0$ \cite{Israel} which require $[h_{MN}]=0$
and
\gat$
\label{jc2}
[K_{\mu\nu}]=-\kappa_5^2\left(T^{(br)}_{\mu\nu}-\frac{1}{3}h_{\mu\nu}\,T^{(br)}\right)\,.$
In the above, $K_{MN}$ is the extrinsic curvature of the brane defined by
$K_{MN}\equiv h^{A}_{\,M} h^{B}_{\,N} \nabla_A n_{B}$, while the bracket notation for a
quantity $X$ simply means 
\eq$\label{bracket-def}
[X]=\lim_{y\ra 0^+}X-\lim_{y\ra 0^-}X=X^{(+)}-X^{(-)}\,.$
The induced line-element on the brane is found to be
\begin{equation}\label{metric-near-br}
ds^2=-\left(1-\frac{2M}{r}\right)dt^2 +\left(1-\frac{2M}{r}\right)^{-1}dr^2+r^2d\Omega^2_2\,,\nonum
\end{equation}
which readily satisfies Israel's first condition. We can now easily determine the components of 
the extrinsic  curvature close to the 3-brane which are found to be
\eq$\label{extr-curv}
K_{MN}=-k \frac{d|y|}{dy}\,\del^{\mu}_M\del^{\nu}_N\, h_{\mu\nu},$
with its trace given by $K=-4k\,(d|y|/dy)$.
Equation \eqref{jc2} may be alternatively written as
\eq$\label{jc2-new}
T^{(br)}_{\mu\nu}=-\frac{1}{\kappa_5^2}\left([K_{\mu\nu}]-h_{\mu\nu}[K]\right)\,.$
Using Eq. (\ref{bracket-def}), the assumed $\mathbf{Z}_2$-symmetry of the model in the bulk
and the components of $K_{\mu\nu}$, we readily find 
\eq$\label{brane-ene-comp}
T^{(br)}_{\mu\nu}=-\frac{6k}{\kappa_5^2}\,h_{\mu\nu}\,.$
Comparing Eqs. \eqref{brane-ene-dec} and \eqref{brane-ene-comp} we easily deduce that
$\sigma=6k/\kappa_5^2>0$, while $\tau_{\mu\nu}=0$. Therefore, no additional distribution of
matter is necessary to be introduced on the brane for its consistent embedding in the bulk. 

One may also derive the effective field equations on the brane---a detailed analysis was
presented in \cite{SMS}. Decomposing  the 5-dimensional Riemann and Ricci
tensors in terms of bulk and brane contributions, one arrives at the following field equations
on the brane
\begin{eqnarray}
\hspace*{-0.2cm}
\bar{G}_{\mu\nu}&=&\frac{2\kappa_5^2}{3}\left[T^{(B)}_{\mu\nu}+\left(T^{(B)}_{yy}-\frac{T^{(B)}}{4}\right)h_{\mu\nu}\right]_{y\ra 0}\nonumber \\[1mm]
&-&3k^2\, h_{\mu\nu} + 8\pi G_4\,\tau_{\mu\nu}+\kappa_5^4\,\pi_{\mu\nu}-E_{\mu\nu}\Big|_{y\ra 0}\,.
\label{grav-eqs-br}
\end{eqnarray}
In the above, $\bar{G}_{\mu\nu}$ denotes  the Einstein tensor on the brane and $G_4=\kappa_5^4\, \sig/48\pi>0$
is the four-dimensional Newton's constant. The quantity $E_{\mu\nu}$ is defined  via the relation
$E_{MN}=C^A{}_{BCD}\, n_A\,n^C\,h^B{}_M\,h^{D}{}_N$ in terms of the  Weyl tensor $C^A{}_{BCD}$,
and $\pi_{\mu\nu}$ is a quadratic function of $\tau_{\mu\nu}$ \cite{SMS}. In our case, $\tau_{\mu\nu}=0$, therefore
$\pi_{\mu\nu}=0$. The components of the quantity inside the square brackets in Eq. (\ref{grav-eqs-br}) involving the
bulk energy-momentum tensor $T ^{(B)}_{\mu\nu}$ may be evaluated to be 
\eq$\label{new-brane-ene-comp1}
\frac{9k^2}{2\kappa_5^2}h_{\mu\nu}
+\frac{3M}{2\kappa_5^2}\frac{1}{r^3}\left(
\begin{array}{cccc}
 -h_{tt} & 0 & 0 & 0  \\
 0 & -h_{rr} & 0 & 0  \\
 0 & 0 & h_{\theta\theta} & 0  \\
 0 & 0 & 0 & h_{\varphi \varphi}
\end{array}
\right)\,,\nonum$

while, employing the 5-dimensional line-element (\ref{metr-r-y}), we also find
\eq$\label{new-brane-ene-comp2}
(E_{\mu\nu})\Big|_{y\ra 0}=
\frac{M}{r^3}\left(
\begin{array}{cccc}
 -h_{tt} & 0 & 0 & 0  \\
 0 & -h_{rr} & 0 & 0  \\
 0 & 0 & h_{\theta\theta} & 0  \\
 0 & 0 & 0 & h_{\varphi \varphi}
\end{array}
\right)\,.\nonum$
Substituting the above two results into Eq. (\ref{grav-eqs-br}), we readily obtain $\bar{G}_{\mu\nu}=0$.
This is indeed the anticipated result since the induced line-element on the brane is described by the
Schwarzschild solution which is a vacuum solution.

\section{CONCLUSIONS}

In this work, we have successfully constructed from first principles the geometry of an analytic
five-dimensional black hole exponentially localized close to our 3-brane. We have demonstrated that
the black-hole singularity lies entirely on the brane, while the event horizon extends into the bulk
but is exponentially suppressed as we move along the extra dimension. This exponential localization
alters the shape of the event horizon, making it  appear as a five-dimensional pancake. The
5-dimensional line-element is effectively AdS$_5$ outside the event horizon and
reduces  to the Schwarzschild solution on the brane. 

The derived geometry is supported by an anisotropic fluid in the bulk described by a diagonal
energy-momentum tensor with only two independent components: the energy density $\rho_E$
and tangential pressure $p_2$. All energy conditions are satisfied on the brane whereas a 
local violation takes place in the bulk in the region inside the event horizon. No additional
matter needs to be introduced on the brane for its consistent embedding in the bulk geometry
while the effective field equations are shown to be satisfied by the vacuum Schwarzschild
geometry on the brane.  

We should note here that a similar perspective for the construction of
the bulk geometry was adopted in \cite{Dai}, however, the form of the 5-dimensional line-element
and bulk energy-momentum tensor did not support either a Schwarzschild solution on the brane or
an AdS$_5$ spacetime right outside the black-hole horizon. Apart from the aforementioned features,
our solution supports an
exponentially decreasing warp factor in the bulk, therefore successfully incorporates
the original Randall-Sundrum brane-world model \cite{RS1,RS2}. Due to this
behavior, our results could be considered also in the context of holography 
\cite{Maldacena,  Gubser, Witten}. In the asymptotic regime where the spacetime becomes
purely AdS$_5$, a four-dimensional conformal field theory (CFT) can be mapped. As we deviate
from the AdS$_5$ limit, the modification in the 5-dimensional metric can be attributed to matter
added in the boundary CFT and related to interesting field-theory phenomena such as chiral
symmetry breaking \cite{Pomarol, Alho}, confinement/deconfinement \cite{Ballon}, etc.

{\bf Acknowledgements.} T.N. is grateful to Thomas Pappas and Konstantinos Rigatos
for useful discussions. The research of T.N. was co-financed by Greece and the European Union 
(European Social Fund- ESF) through the Operational Programme “Human Resources Development, 
Education and Lifelong Learning” in the context of the project “Strengthening Human Resources 
Research Potential via Doctorate Research – 2nd Cycle” (MIS-5000432), implemented by the State 
Scholarships Foundation (IKY).\vspace{-0.5em}

\appendix

\section{Appendix A: Scalar Curvature Quantities}
\label{Sc-Curv}

In terms of the $(\rho,\chi)$-coordinates, the curvature invariants $\mathcal{R}\equiv R_{MN}\,R^{MN}$ and $\mathcal{K}\equiv R_{MNKL}\,R^{MNKL}$ are given by
\bal$
\mathcal{R}&=80 k^4-\frac{96 k^4 M \cos^2\chi}{\rho }+\frac{24 k^3 M \cos \chi \left(3 k M \cos ^3\chi+8\right)}{\rho ^2}\nonum\\[1mm]
&\hsp -\frac{8 k^2 M \left(9 k M \cos ^3\chi+4\right)}{\rho ^3}+\frac{102 k^2 M^2 \cos ^2\chi }{\rho ^4}\nonum\\[1mm]
&\hsp-\frac{60 k M^2 \cos \chi }{\rho ^5}+\frac{14 M^2}{\rho ^6}\,,$
\bal$\mathcal{K}&=40 k^4-\frac{48 k^4 M \cos ^2(\chi )}{\rho }+\frac{48 k^3 M \cos \chi  \left(3 k M \cos ^3\chi +2\right)}{\rho ^2}\nonum\\[1mm]
&\hsp+\frac{8 k^2 M \left(36 k M \cos ^3\chi -2\right)}{\rho ^3}+\frac{456 k^2 M^2 \cos ^2\chi }{\rho ^4}\nonum\\[1mm]
&\hsp +\frac{240 k M^2 \cos \chi }{\rho ^5}+\frac{88 M^2}{\rho^6}\,,$
while in terms of $(r,y)$-coordinates, we get
\bal$
\mathcal{R}&=\frac{2k^6M^2 \left(160-384\, e^{k \left|y\right| }+375\, e^{2 k \left| y\right| }-180\, e^{3 k \left| y\right| }+36
\, e^{4 k \left| y\right| }\right)}{\left[k^2r^2+\left(e^{k \left| y\right| }-1\right)^2\right]^3}\nonum\\[2mm]
&\hsp +80k^4-\frac{32 k^5 M \left(10-12\, e^{k \left| y\right| }+3\, e^{2 k \left| y\right| }\right)}{\left[k^2r^2+\left(e^{k \left| y
\right| }-1\right)^2\right]^{3/2}}\,,$
\bal$\mathcal{K}&=\frac{8k^6M^2 \left(20-48\, e^{k \left|y\right| }+57\, e^{2 k \left| y\right| }-36\, e^{3 k \left| y\right| }+18
\, e^{4 k \left| y\right| }\right)}{\left[k^2r^2+\left(e^{k \left| y\right| }-1\right)^2\right]^3}\nonum\\[2mm]
&\hsp +40k^4-\frac{16 k^5 M \left(10-12\, e^{k \left| y\right| }+3\, e^{2 k \left| y\right| }\right)}{\left[k^2r^2+\left(e^{k \left| y\right| }-1\right)^2\right]^{3/2}}\,.$

\bibliography{Bibliography}{}

\begin{thebibliography}{59}%
\makeatletter
\providecommand \@ifxundefined [1]{%
 \@ifx{#1\undefined}
}%
\providecommand \@ifnum [1]{%
 \ifnum #1\expandafter \@firstoftwo
 \else \expandafter \@secondoftwo
 \fi
}%
\providecommand \@ifx [1]{%
 \ifx #1\expandafter \@firstoftwo
 \else \expandafter \@secondoftwo
 \fi
}%
\providecommand \natexlab [1]{#1}%
\providecommand \enquote  [1]{``#1''}%
\providecommand \bibnamefont  [1]{#1}%
\providecommand \bibfnamefont [1]{#1}%
\providecommand \citenamefont [1]{#1}%
\providecommand \href@noop [0]{\@secondoftwo}%
\providecommand \href [0]{\begingroup \@sanitize@url \@href}%
\providecommand \@href[1]{\@@startlink{#1}\@@href}%
\providecommand \@@href[1]{\endgroup#1\@@endlink}%
\providecommand \@sanitize@url [0]{\catcode `\\12\catcode `\$12\catcode
  `\&12\catcode `\#12\catcode `\^12\catcode `\_12\catcode `\%12\relax}%
\providecommand \@@startlink[1]{}%
\providecommand \@@endlink[0]{}%
\providecommand \url  [0]{\begingroup\@sanitize@url \@url }%
\providecommand \@url [1]{\endgroup\@href {#1}{\urlprefix }}%
\providecommand \urlprefix  [0]{URL }%
\providecommand \Eprint [0]{\href }%
\providecommand \doibase [0]{http://dx.doi.org/}%
\providecommand \selectlanguage [0]{\@gobble}%
\providecommand \bibinfo  [0]{\@secondoftwo}%
\providecommand \bibfield  [0]{\@secondoftwo}%
\providecommand \translation [1]{[#1]}%
\providecommand \BibitemOpen [0]{}%
\providecommand \bibitemStop [0]{}%
\providecommand \bibitemNoStop [0]{.\EOS\space}%
\providecommand \EOS [0]{\spacefactor3000\relax}%
\providecommand \BibitemShut  [1]{\csname bibitem#1\endcsname}%
\let\auto@bib@innerbib\@empty
\bibitem [{\citenamefont {Randall}\ and\ \citenamefont
  {Sundrum}(1999{\natexlab{a}})}]{RS1}%
  \BibitemOpen
  \bibfield  {author} {\bibinfo {author} {\bibfnamefont {L.}~\bibnamefont
  {Randall}}\ and\ \bibinfo {author} {\bibfnamefont {R.}~\bibnamefont
  {Sundrum}},\ }\href {\doibase 10.1103/PhysRevLett.83.3370} {\bibfield
  {journal} {\bibinfo  {journal} {Phys. Rev. Lett.}\ }\textbf {\bibinfo
  {volume} {83}},\ \bibinfo {pages} {3370} (\bibinfo {year}
  {1999}{\natexlab{a}})},\ \Eprint {http://arxiv.org/abs/hep-ph/9905221}
  {arXiv:hep-ph/9905221} \BibitemShut {NoStop}%
\bibitem [{\citenamefont {Randall}\ and\ \citenamefont
  {Sundrum}(1999{\natexlab{b}})}]{RS2}%
  \BibitemOpen
  \bibfield  {author} {\bibinfo {author} {\bibfnamefont {L.}~\bibnamefont
  {Randall}}\ and\ \bibinfo {author} {\bibfnamefont {R.}~\bibnamefont
  {Sundrum}},\ }\href {\doibase 10.1103/PhysRevLett.83.4690} {\bibfield
  {journal} {\bibinfo  {journal} {Phys. Rev. Lett.}\ }\textbf {\bibinfo
  {volume} {83}},\ \bibinfo {pages} {4690} (\bibinfo {year}
  {1999}{\natexlab{b}})},\ \Eprint {http://arxiv.org/abs/hep-th/9906064}
  {arXiv:hep-th/9906064} \BibitemShut {NoStop}%
\bibitem [{\citenamefont {Rubakov}\ and\ \citenamefont
  {Shaposhnikov}(1983)}]{misha}%
  \BibitemOpen
  \bibfield  {author} {\bibinfo {author} {\bibfnamefont {V.~A.}\ \bibnamefont
  {Rubakov}}\ and\ \bibinfo {author} {\bibfnamefont {M.~E.}\ \bibnamefont
  {Shaposhnikov}},\ }\href {\doibase 10.1016/0370-2693(83)91253-4} {\bibfield
  {journal} {\bibinfo  {journal} {Phys. Lett. B}\ }\textbf {\bibinfo {volume}
  {125}},\ \bibinfo {pages} {136} (\bibinfo {year} {1983})}\BibitemShut
  {NoStop}%
\bibitem [{\citenamefont {Akama}(1982)}]{akama}%
  \BibitemOpen
  \bibfield  {author} {\bibinfo {author} {\bibfnamefont {K.}~\bibnamefont
  {Akama}},\ }\href@noop {} {\bibfield  {journal} {\bibinfo  {journal} {Lect.
  Notes Phys.}\ }\textbf {\bibinfo {volume} {176}},\ \bibinfo {pages} {267}
  (\bibinfo {year} {1982})},\ \Eprint {http://arxiv.org/abs/hep-th/0001113}
  {arXiv:hep-th/0001113} \BibitemShut {NoStop}%
\bibitem [{\citenamefont {Chamblin}\ \emph {et~al.}(2000)\citenamefont
  {Chamblin}, \citenamefont {Hawking},\ and\ \citenamefont {Reall}}]{CHR}%
  \BibitemOpen
  \bibfield  {author} {\bibinfo {author} {\bibfnamefont {A.}~\bibnamefont
  {Chamblin}}, \bibinfo {author} {\bibfnamefont {S.}~\bibnamefont {Hawking}}, \
  and\ \bibinfo {author} {\bibfnamefont {H.}~\bibnamefont {Reall}},\ }\href
  {\doibase 10.1103/PhysRevD.61.065007} {\bibfield  {journal} {\bibinfo
  {journal} {Phys. Rev. D}\ }\textbf {\bibinfo {volume} {61}},\ \bibinfo
  {pages} {065007} (\bibinfo {year} {2000})},\ \Eprint
  {http://arxiv.org/abs/hep-th/9909205} {arXiv:hep-th/9909205} \BibitemShut
  {NoStop}%
\bibitem [{\citenamefont {Gregory}\ and\ \citenamefont {Laflamme}(1993)}]{GL}%
  \BibitemOpen
  \bibfield  {author} {\bibinfo {author} {\bibfnamefont {R.}~\bibnamefont
  {Gregory}}\ and\ \bibinfo {author} {\bibfnamefont {R.}~\bibnamefont
  {Laflamme}},\ }\href {\doibase 10.1103/PhysRevLett.70.2837} {\bibfield
  {journal} {\bibinfo  {journal} {Phys. Rev. Lett.}\ }\textbf {\bibinfo
  {volume} {70}},\ \bibinfo {pages} {2837} (\bibinfo {year} {1993})},\ \Eprint
  {http://arxiv.org/abs/hep-th/9301052} {arXiv:hep-th/9301052} \BibitemShut
  {NoStop}%
\bibitem [{\citenamefont {Gregory}(2000)}]{Ruth}%
  \BibitemOpen
  \bibfield  {author} {\bibinfo {author} {\bibfnamefont {R.}~\bibnamefont
  {Gregory}},\ }\href {\doibase 10.1088/0264-9381/17/18/103} {\bibfield
  {journal} {\bibinfo  {journal} {Class. Quant. Grav.}\ }\textbf {\bibinfo
  {volume} {17}},\ \bibinfo {pages} {L125} (\bibinfo {year} {2000})},\ \Eprint
  {http://arxiv.org/abs/hep-th/0004101} {arXiv:hep-th/0004101} \BibitemShut
  {NoStop}%
\bibitem [{\citenamefont {Kudoh}\ \emph {et~al.}(2003)\citenamefont {Kudoh},
  \citenamefont {Tanaka},\ and\ \citenamefont {Nakamura}}]{Kudoh1}%
  \BibitemOpen
  \bibfield  {author} {\bibinfo {author} {\bibfnamefont {H.}~\bibnamefont
  {Kudoh}}, \bibinfo {author} {\bibfnamefont {T.}~\bibnamefont {Tanaka}}, \
  and\ \bibinfo {author} {\bibfnamefont {T.}~\bibnamefont {Nakamura}},\ }\href
  {\doibase 10.1103/PhysRevD.68.024035} {\bibfield  {journal} {\bibinfo
  {journal} {Phys. Rev. D}\ }\textbf {\bibinfo {volume} {68}},\ \bibinfo
  {pages} {024035} (\bibinfo {year} {2003})},\ \Eprint
  {http://arxiv.org/abs/gr-qc/0301089} {arXiv:gr-qc/0301089} \BibitemShut
  {NoStop}%
\bibitem [{\citenamefont {Kudoh}(2004)}]{Kudoh2}%
  \BibitemOpen
  \bibfield  {author} {\bibinfo {author} {\bibfnamefont {H.}~\bibnamefont
  {Kudoh}},\ }\href {\doibase 10.1103/PhysRevD.70.029901} {\bibfield  {journal}
  {\bibinfo  {journal} {Phys. Rev. D}\ }\textbf {\bibinfo {volume} {69}},\
  \bibinfo {pages} {104019} (\bibinfo {year} {2004})},\ \bibinfo {note}
  {[Erratum: Phys.Rev.D 70, 029901 (2004)]},\ \Eprint
  {http://arxiv.org/abs/hep-th/0401229} {arXiv:hep-th/0401229} \BibitemShut
  {NoStop}%
\bibitem [{\citenamefont {Tanahashi}\ and\ \citenamefont
  {Tanaka}(2008)}]{Tanahashi}%
  \BibitemOpen
  \bibfield  {author} {\bibinfo {author} {\bibfnamefont {N.}~\bibnamefont
  {Tanahashi}}\ and\ \bibinfo {author} {\bibfnamefont {T.}~\bibnamefont
  {Tanaka}},\ }\href {\doibase 10.1088/1126-6708/2008/03/041} {\bibfield
  {journal} {\bibinfo  {journal} {JHEP}\ }\textbf {\bibinfo {volume} {03}},\
  \bibinfo {pages} {041} (\bibinfo {year} {2008})},\ \Eprint
  {http://arxiv.org/abs/0712.3799} {arXiv:0712.3799 [gr-qc]} \BibitemShut
  {NoStop}%
\bibitem [{\citenamefont {Figueras}\ and\ \citenamefont
  {Wiseman}(2011)}]{Figueras}%
  \BibitemOpen
  \bibfield  {author} {\bibinfo {author} {\bibfnamefont {P.}~\bibnamefont
  {Figueras}}\ and\ \bibinfo {author} {\bibfnamefont {T.}~\bibnamefont
  {Wiseman}},\ }\href {\doibase 10.1103/PhysRevLett.107.081101} {\bibfield
  {journal} {\bibinfo  {journal} {Phys. Rev. Lett.}\ }\textbf {\bibinfo
  {volume} {107}},\ \bibinfo {pages} {081101} (\bibinfo {year} {2011})},\
  \Eprint {http://arxiv.org/abs/1105.2558} {arXiv:1105.2558 [hep-th]}
  \BibitemShut {NoStop}%
\bibitem [{\citenamefont {Abdolrahimi}\ \emph {et~al.}(2013)\citenamefont
  {Abdolrahimi}, \citenamefont {Cattoen}, \citenamefont {Page},\ and\
  \citenamefont {Yaghoobpour-Tari}}]{Page}%
  \BibitemOpen
  \bibfield  {author} {\bibinfo {author} {\bibfnamefont {S.}~\bibnamefont
  {Abdolrahimi}}, \bibinfo {author} {\bibfnamefont {C.}~\bibnamefont
  {Cattoen}}, \bibinfo {author} {\bibfnamefont {D.~N.}\ \bibnamefont {Page}}, \
  and\ \bibinfo {author} {\bibfnamefont {S.}~\bibnamefont {Yaghoobpour-Tari}},\
  }\href {\doibase 10.1016/j.physletb.2013.02.034} {\bibfield  {journal}
  {\bibinfo  {journal} {Phys. Lett. B}\ }\textbf {\bibinfo {volume} {720}},\
  \bibinfo {pages} {405} (\bibinfo {year} {2013})},\ \Eprint
  {http://arxiv.org/abs/1206.0708} {arXiv:1206.0708 [hep-th]} \BibitemShut
  {NoStop}%
\bibitem [{\citenamefont {Emparan}\ \emph {et~al.}(2000)\citenamefont
  {Emparan}, \citenamefont {Horowitz},\ and\ \citenamefont {Myers}}]{EHM}%
  \BibitemOpen
  \bibfield  {author} {\bibinfo {author} {\bibfnamefont {R.}~\bibnamefont
  {Emparan}}, \bibinfo {author} {\bibfnamefont {G.~T.}\ \bibnamefont
  {Horowitz}}, \ and\ \bibinfo {author} {\bibfnamefont {R.~C.}\ \bibnamefont
  {Myers}},\ }\href {\doibase 10.1088/1126-6708/2000/01/007} {\bibfield
  {journal} {\bibinfo  {journal} {JHEP}\ }\textbf {\bibinfo {volume} {01}},\
  \bibinfo {pages} {007} (\bibinfo {year} {2000})},\ \Eprint
  {http://arxiv.org/abs/hep-th/9911043} {arXiv:hep-th/9911043} \BibitemShut
  {NoStop}%
\bibitem [{\citenamefont {Dadhich}(2000)}]{Dadhich}%
  \BibitemOpen
  \bibfield  {author} {\bibinfo {author} {\bibfnamefont {N.}~\bibnamefont
  {Dadhich}},\ }\href {\doibase 10.1016/S0370-2693(00)01101-1} {\bibfield
  {journal} {\bibinfo  {journal} {Phys. Lett. B}\ }\textbf {\bibinfo {volume}
  {492}},\ \bibinfo {pages} {357} (\bibinfo {year} {2000})},\ \Eprint
  {http://arxiv.org/abs/hep-th/0009178} {arXiv:hep-th/0009178} \BibitemShut
  {NoStop}%
\bibitem [{\citenamefont {Dadhich}\ \emph {et~al.}(2000)\citenamefont
  {Dadhich}, \citenamefont {Maartens}, \citenamefont {Papadopoulos},\ and\
  \citenamefont {Rezania}}]{tidal}%
  \BibitemOpen
  \bibfield  {author} {\bibinfo {author} {\bibfnamefont {N.}~\bibnamefont
  {Dadhich}}, \bibinfo {author} {\bibfnamefont {R.}~\bibnamefont {Maartens}},
  \bibinfo {author} {\bibfnamefont {P.}~\bibnamefont {Papadopoulos}}, \ and\
  \bibinfo {author} {\bibfnamefont {V.}~\bibnamefont {Rezania}},\ }\href
  {\doibase 10.1016/S0370-2693(00)00798-X} {\bibfield  {journal} {\bibinfo
  {journal} {Phys. Lett. B}\ }\textbf {\bibinfo {volume} {487}},\ \bibinfo
  {pages} {1} (\bibinfo {year} {2000})},\ \Eprint
  {http://arxiv.org/abs/hep-th/0003061} {arXiv:hep-th/0003061} \BibitemShut
  {NoStop}%
\bibitem [{\citenamefont {Kofinas}\ \emph {et~al.}(2002)\citenamefont
  {Kofinas}, \citenamefont {Papantonopoulos},\ and\ \citenamefont
  {Zamarias}}]{Papanto}%
  \BibitemOpen
  \bibfield  {author} {\bibinfo {author} {\bibfnamefont {G.}~\bibnamefont
  {Kofinas}}, \bibinfo {author} {\bibfnamefont {E.}~\bibnamefont
  {Papantonopoulos}}, \ and\ \bibinfo {author} {\bibfnamefont {V.}~\bibnamefont
  {Zamarias}},\ }\href {\doibase 10.1103/PhysRevD.66.104028} {\bibfield
  {journal} {\bibinfo  {journal} {Phys. Rev. D}\ }\textbf {\bibinfo {volume}
  {66}},\ \bibinfo {pages} {104028} (\bibinfo {year} {2002})},\ \Eprint
  {http://arxiv.org/abs/hep-th/0208207} {arXiv:hep-th/0208207} \BibitemShut
  {NoStop}%
\bibitem [{\citenamefont {Bruni}\ \emph {et~al.}(2001)\citenamefont {Bruni},
  \citenamefont {Germani},\ and\ \citenamefont {Maartens}}]{Bruni}%
  \BibitemOpen
  \bibfield  {author} {\bibinfo {author} {\bibfnamefont {M.}~\bibnamefont
  {Bruni}}, \bibinfo {author} {\bibfnamefont {C.}~\bibnamefont {Germani}}, \
  and\ \bibinfo {author} {\bibfnamefont {R.}~\bibnamefont {Maartens}},\ }\href
  {\doibase 10.1103/PhysRevLett.87.231302} {\bibfield  {journal} {\bibinfo
  {journal} {Phys. Rev. Lett.}\ }\textbf {\bibinfo {volume} {87}},\ \bibinfo
  {pages} {231302} (\bibinfo {year} {2001})},\ \Eprint
  {http://arxiv.org/abs/gr-qc/0108013} {arXiv:gr-qc/0108013} \BibitemShut
  {NoStop}%
\bibitem [{\citenamefont {Kanti}\ and\ \citenamefont {Tamvakis}(2002)}]{KT}%
  \BibitemOpen
  \bibfield  {author} {\bibinfo {author} {\bibfnamefont {P.}~\bibnamefont
  {Kanti}}\ and\ \bibinfo {author} {\bibfnamefont {K.}~\bibnamefont
  {Tamvakis}},\ }\href {\doibase 10.1103/PhysRevD.65.084010} {\bibfield
  {journal} {\bibinfo  {journal} {Phys. Rev. D}\ }\textbf {\bibinfo {volume}
  {65}},\ \bibinfo {pages} {084010} (\bibinfo {year} {2002})},\ \Eprint
  {http://arxiv.org/abs/hep-th/0110298} {arXiv:hep-th/0110298} \BibitemShut
  {NoStop}%
\bibitem [{\citenamefont {Govender}\ and\ \citenamefont
  {Dadhich}(2002)}]{Dadhich2}%
  \BibitemOpen
  \bibfield  {author} {\bibinfo {author} {\bibfnamefont {M.}~\bibnamefont
  {Govender}}\ and\ \bibinfo {author} {\bibfnamefont {N.}~\bibnamefont
  {Dadhich}},\ }\href {\doibase 10.1016/S0370-2693(02)01996-2} {\bibfield
  {journal} {\bibinfo  {journal} {Phys. Lett. B}\ }\textbf {\bibinfo {volume}
  {538}},\ \bibinfo {pages} {233} (\bibinfo {year} {2002})},\ \Eprint
  {http://arxiv.org/abs/hep-th/0109086} {arXiv:hep-th/0109086} \BibitemShut
  {NoStop}%
\bibitem [{\citenamefont {Casadio}\ \emph {et~al.}(2002)\citenamefont
  {Casadio}, \citenamefont {Fabbri},\ and\ \citenamefont
  {Mazzacurati}}]{CasadioNew}%
  \BibitemOpen
  \bibfield  {author} {\bibinfo {author} {\bibfnamefont {R.}~\bibnamefont
  {Casadio}}, \bibinfo {author} {\bibfnamefont {A.}~\bibnamefont {Fabbri}}, \
  and\ \bibinfo {author} {\bibfnamefont {L.}~\bibnamefont {Mazzacurati}},\
  }\href {\doibase 10.1103/PhysRevD.65.084040} {\bibfield  {journal} {\bibinfo
  {journal} {Phys. Rev. D}\ }\textbf {\bibinfo {volume} {65}},\ \bibinfo
  {pages} {084040} (\bibinfo {year} {2002})},\ \Eprint
  {http://arxiv.org/abs/gr-qc/0111072} {arXiv:gr-qc/0111072} \BibitemShut
  {NoStop}%
\bibitem [{\citenamefont {Emparan}\ \emph {et~al.}(2002)\citenamefont
  {Emparan}, \citenamefont {Fabbri},\ and\ \citenamefont {Kaloper}}]{EFK}%
  \BibitemOpen
  \bibfield  {author} {\bibinfo {author} {\bibfnamefont {R.}~\bibnamefont
  {Emparan}}, \bibinfo {author} {\bibfnamefont {A.}~\bibnamefont {Fabbri}}, \
  and\ \bibinfo {author} {\bibfnamefont {N.}~\bibnamefont {Kaloper}},\ }\href
  {\doibase 10.1088/1126-6708/2002/08/043} {\bibfield  {journal} {\bibinfo
  {journal} {JHEP}\ }\textbf {\bibinfo {volume} {08}},\ \bibinfo {pages} {043}
  (\bibinfo {year} {2002})},\ \Eprint {http://arxiv.org/abs/hep-th/0206155}
  {arXiv:hep-th/0206155} \BibitemShut {NoStop}%
\bibitem [{\citenamefont {Frolov}\ \emph {et~al.}(2003)\citenamefont {Frolov},
  \citenamefont {Snajdr},\ and\ \citenamefont {Stojkovic}}]{Frolov}%
  \BibitemOpen
  \bibfield  {author} {\bibinfo {author} {\bibfnamefont {V.~P.}\ \bibnamefont
  {Frolov}}, \bibinfo {author} {\bibfnamefont {M.}~\bibnamefont {Snajdr}}, \
  and\ \bibinfo {author} {\bibfnamefont {D.}~\bibnamefont {Stojkovic}},\ }\href
  {\doibase 10.1103/PhysRevD.68.044002} {\bibfield  {journal} {\bibinfo
  {journal} {Phys. Rev. D}\ }\textbf {\bibinfo {volume} {68}},\ \bibinfo
  {pages} {044002} (\bibinfo {year} {2003})},\ \Eprint
  {http://arxiv.org/abs/gr-qc/0304083} {arXiv:gr-qc/0304083} \BibitemShut
  {NoStop}%
\bibitem [{\citenamefont {Emparan}\ \emph {et~al.}(2003)\citenamefont
  {Emparan}, \citenamefont {Garcia-Bellido},\ and\ \citenamefont
  {Kaloper}}]{EGK}%
  \BibitemOpen
  \bibfield  {author} {\bibinfo {author} {\bibfnamefont {R.}~\bibnamefont
  {Emparan}}, \bibinfo {author} {\bibfnamefont {J.}~\bibnamefont
  {Garcia-Bellido}}, \ and\ \bibinfo {author} {\bibfnamefont {N.}~\bibnamefont
  {Kaloper}},\ }\href {\doibase 10.1088/1126-6708/2003/01/079} {\bibfield
  {journal} {\bibinfo  {journal} {JHEP}\ }\textbf {\bibinfo {volume} {01}},\
  \bibinfo {pages} {079} (\bibinfo {year} {2003})},\ \Eprint
  {http://arxiv.org/abs/hep-th/0212132} {arXiv:hep-th/0212132} \BibitemShut
  {NoStop}%
\bibitem [{\citenamefont {Tanaka}(2003)}]{Tanaka}%
  \BibitemOpen
  \bibfield  {author} {\bibinfo {author} {\bibfnamefont {T.}~\bibnamefont
  {Tanaka}},\ }\href {\doibase 10.1143/PTPS.148.307} {\bibfield  {journal}
  {\bibinfo  {journal} {Prog. Theor. Phys. Suppl.}\ }\textbf {\bibinfo {volume}
  {148}},\ \bibinfo {pages} {307} (\bibinfo {year} {2003})},\ \Eprint
  {http://arxiv.org/abs/gr-qc/0203082} {arXiv:gr-qc/0203082} \BibitemShut
  {NoStop}%
\bibitem [{\citenamefont {Kanti}\ \emph {et~al.}(2003)\citenamefont {Kanti},
  \citenamefont {Olasagasti},\ and\ \citenamefont {Tamvakis}}]{KOT}%
  \BibitemOpen
  \bibfield  {author} {\bibinfo {author} {\bibfnamefont {P.}~\bibnamefont
  {Kanti}}, \bibinfo {author} {\bibfnamefont {I.}~\bibnamefont {Olasagasti}}, \
  and\ \bibinfo {author} {\bibfnamefont {K.}~\bibnamefont {Tamvakis}},\ }\href
  {\doibase 10.1103/PhysRevD.68.124001} {\bibfield  {journal} {\bibinfo
  {journal} {Phys. Rev. D}\ }\textbf {\bibinfo {volume} {68}},\ \bibinfo
  {pages} {124001} (\bibinfo {year} {2003})},\ \Eprint
  {http://arxiv.org/abs/hep-th/0307201} {arXiv:hep-th/0307201} \BibitemShut
  {NoStop}%
\bibitem [{\citenamefont {Charmousis}\ and\ \citenamefont
  {Gregory}(2004)}]{Charmousis}%
  \BibitemOpen
  \bibfield  {author} {\bibinfo {author} {\bibfnamefont {C.}~\bibnamefont
  {Charmousis}}\ and\ \bibinfo {author} {\bibfnamefont {R.}~\bibnamefont
  {Gregory}},\ }\href {\doibase 10.1088/0264-9381/21/2/016} {\bibfield
  {journal} {\bibinfo  {journal} {Class. Quant. Grav.}\ }\textbf {\bibinfo
  {volume} {21}},\ \bibinfo {pages} {527} (\bibinfo {year} {2004})},\ \Eprint
  {http://arxiv.org/abs/gr-qc/0306069} {arXiv:gr-qc/0306069} \BibitemShut
  {NoStop}%
\bibitem [{\citenamefont {Kofinas}\ and\ \citenamefont
  {Papantonopoulos}(2004)}]{Kofinas}%
  \BibitemOpen
  \bibfield  {author} {\bibinfo {author} {\bibfnamefont {G.}~\bibnamefont
  {Kofinas}}\ and\ \bibinfo {author} {\bibfnamefont {E.}~\bibnamefont
  {Papantonopoulos}},\ }\href {\doibase 10.1088/1475-7516/2004/12/011}
  {\bibfield  {journal} {\bibinfo  {journal} {JCAP}\ }\textbf {\bibinfo
  {volume} {12}},\ \bibinfo {pages} {011} (\bibinfo {year} {2004})},\ \Eprint
  {http://arxiv.org/abs/gr-qc/0401047} {arXiv:gr-qc/0401047} \BibitemShut
  {NoStop}%
\bibitem [{\citenamefont {Shankaranarayanan}\ and\ \citenamefont
  {Dadhich}(2004)}]{Shanka}%
  \BibitemOpen
  \bibfield  {author} {\bibinfo {author} {\bibfnamefont {S.}~\bibnamefont
  {Shankaranarayanan}}\ and\ \bibinfo {author} {\bibfnamefont {N.}~\bibnamefont
  {Dadhich}},\ }\href {\doibase 10.1142/S0218271804005109} {\bibfield
  {journal} {\bibinfo  {journal} {Int. J. Mod. Phys. D}\ }\textbf {\bibinfo
  {volume} {13}},\ \bibinfo {pages} {1095} (\bibinfo {year} {2004})},\ \Eprint
  {http://arxiv.org/abs/gr-qc/0306111} {arXiv:gr-qc/0306111} \BibitemShut
  {NoStop}%
\bibitem [{\citenamefont {Karasik}\ \emph {et~al.}(2004)\citenamefont
  {Karasik}, \citenamefont {Sahabandu}, \citenamefont {Suranyi},\ and\
  \citenamefont {Wijewardhana}}]{Karasik}%
  \BibitemOpen
  \bibfield  {author} {\bibinfo {author} {\bibfnamefont {D.}~\bibnamefont
  {Karasik}}, \bibinfo {author} {\bibfnamefont {C.}~\bibnamefont {Sahabandu}},
  \bibinfo {author} {\bibfnamefont {P.}~\bibnamefont {Suranyi}}, \ and\
  \bibinfo {author} {\bibfnamefont {L.~C.~R.}\ \bibnamefont {Wijewardhana}},\
  }\href {\doibase 10.1103/PhysRevD.70.064007} {\bibfield  {journal} {\bibinfo
  {journal} {Phys. Rev. D}\ }\textbf {\bibinfo {volume} {70}},\ \bibinfo
  {pages} {064007} (\bibinfo {year} {2004})},\ \Eprint
  {http://arxiv.org/abs/gr-qc/0404015} {arXiv:gr-qc/0404015} \BibitemShut
  {NoStop}%
\bibitem [{\citenamefont {Galfard}\ \emph {et~al.}(2006)\citenamefont
  {Galfard}, \citenamefont {Germani},\ and\ \citenamefont {Ishibashi}}]{GGI}%
  \BibitemOpen
  \bibfield  {author} {\bibinfo {author} {\bibfnamefont {C.}~\bibnamefont
  {Galfard}}, \bibinfo {author} {\bibfnamefont {C.}~\bibnamefont {Germani}}, \
  and\ \bibinfo {author} {\bibfnamefont {A.}~\bibnamefont {Ishibashi}},\ }\href
  {\doibase 10.1103/PhysRevD.73.064014} {\bibfield  {journal} {\bibinfo
  {journal} {Phys. Rev. D}\ }\textbf {\bibinfo {volume} {73}},\ \bibinfo
  {pages} {064014} (\bibinfo {year} {2006})},\ \Eprint
  {http://arxiv.org/abs/hep-th/0512001} {arXiv:hep-th/0512001} \BibitemShut
  {NoStop}%
\bibitem [{\citenamefont {Creek}\ \emph {et~al.}(2006)\citenamefont {Creek},
  \citenamefont {Gregory}, \citenamefont {Kanti},\ and\ \citenamefont
  {Mistry}}]{CGKM}%
  \BibitemOpen
  \bibfield  {author} {\bibinfo {author} {\bibfnamefont {S.}~\bibnamefont
  {Creek}}, \bibinfo {author} {\bibfnamefont {R.}~\bibnamefont {Gregory}},
  \bibinfo {author} {\bibfnamefont {P.}~\bibnamefont {Kanti}}, \ and\ \bibinfo
  {author} {\bibfnamefont {B.}~\bibnamefont {Mistry}},\ }\href {\doibase
  10.1088/0264-9381/23/23/004} {\bibfield  {journal} {\bibinfo  {journal}
  {Class. Quant. Grav.}\ }\textbf {\bibinfo {volume} {23}},\ \bibinfo {pages}
  {6633} (\bibinfo {year} {2006})},\ \Eprint
  {http://arxiv.org/abs/hep-th/0606006} {arXiv:hep-th/0606006} \BibitemShut
  {NoStop}%
\bibitem [{\citenamefont {Casadio}\ and\ \citenamefont
  {Ovalle}(2012)}]{Ovalle}%
  \BibitemOpen
  \bibfield  {author} {\bibinfo {author} {\bibfnamefont {R.}~\bibnamefont
  {Casadio}}\ and\ \bibinfo {author} {\bibfnamefont {J.}~\bibnamefont
  {Ovalle}},\ }\href {\doibase 10.1016/j.physletb.2012.07.041} {\bibfield
  {journal} {\bibinfo  {journal} {Phys. Lett. B}\ }\textbf {\bibinfo {volume}
  {715}},\ \bibinfo {pages} {251} (\bibinfo {year} {2012})},\ \Eprint
  {http://arxiv.org/abs/1201.6145} {arXiv:1201.6145 [gr-qc]} \BibitemShut
  {NoStop}%
\bibitem [{\citenamefont {Harko}\ and\ \citenamefont {Lake}(2014)}]{Harko}%
  \BibitemOpen
  \bibfield  {author} {\bibinfo {author} {\bibfnamefont {T.}~\bibnamefont
  {Harko}}\ and\ \bibinfo {author} {\bibfnamefont {M.~J.}\ \bibnamefont
  {Lake}},\ }\href {\doibase 10.1103/PhysRevD.89.064038} {\bibfield  {journal}
  {\bibinfo  {journal} {Phys. Rev. D}\ }\textbf {\bibinfo {volume} {89}},\
  \bibinfo {pages} {064038} (\bibinfo {year} {2014})},\ \Eprint
  {http://arxiv.org/abs/1312.1420} {arXiv:1312.1420 [gr-qc]} \BibitemShut
  {NoStop}%
\bibitem [{\citenamefont {Herrera-Aguilar}\ \emph {et~al.}(2015)\citenamefont
  {Herrera-Aguilar}, \citenamefont {Kuerten},\ and\ \citenamefont
  {da~Rocha}}]{daRocha1}%
  \BibitemOpen
  \bibfield  {author} {\bibinfo {author} {\bibfnamefont {A.}~\bibnamefont
  {Herrera-Aguilar}}, \bibinfo {author} {\bibfnamefont {A.~M.}\ \bibnamefont
  {Kuerten}}, \ and\ \bibinfo {author} {\bibfnamefont {R.}~\bibnamefont
  {da~Rocha}},\ }\href {\doibase 10.1155/2015/359268} {\bibfield  {journal}
  {\bibinfo  {journal} {Adv. High Energy Phys.}\ }\textbf {\bibinfo {volume}
  {2015}},\ \bibinfo {pages} {359268} (\bibinfo {year} {2015})},\ \Eprint
  {http://arxiv.org/abs/1501.07629} {arXiv:1501.07629 [gr-qc]} \BibitemShut
  {NoStop}%
\bibitem [{\citenamefont {Anber}\ and\ \citenamefont {Sorbo}(2008)}]{AS}%
  \BibitemOpen
  \bibfield  {author} {\bibinfo {author} {\bibfnamefont {M.}~\bibnamefont
  {Anber}}\ and\ \bibinfo {author} {\bibfnamefont {L.}~\bibnamefont {Sorbo}},\
  }\href {\doibase 10.1088/1126-6708/2008/07/098} {\bibfield  {journal}
  {\bibinfo  {journal} {JHEP}\ }\textbf {\bibinfo {volume} {07}},\ \bibinfo
  {pages} {098} (\bibinfo {year} {2008})},\ \Eprint
  {http://arxiv.org/abs/0803.2242} {arXiv:0803.2242 [hep-th]} \BibitemShut
  {NoStop}%
\bibitem [{\citenamefont {Fitzpatrick}\ \emph {et~al.}(2006)\citenamefont
  {Fitzpatrick}, \citenamefont {Randall},\ and\ \citenamefont
  {Wiseman}}]{Fitzpatrick}%
  \BibitemOpen
  \bibfield  {author} {\bibinfo {author} {\bibfnamefont {A.~L.}\ \bibnamefont
  {Fitzpatrick}}, \bibinfo {author} {\bibfnamefont {L.}~\bibnamefont
  {Randall}}, \ and\ \bibinfo {author} {\bibfnamefont {T.}~\bibnamefont
  {Wiseman}},\ }\href {\doibase 10.1088/1126-6708/2006/11/033} {\bibfield
  {journal} {\bibinfo  {journal} {JHEP}\ }\textbf {\bibinfo {volume} {11}},\
  \bibinfo {pages} {033} (\bibinfo {year} {2006})},\ \Eprint
  {http://arxiv.org/abs/hep-th/0608208} {arXiv:hep-th/0608208} \BibitemShut
  {NoStop}%
\bibitem [{\citenamefont {Gregory}\ \emph {et~al.}(2008)\citenamefont
  {Gregory}, \citenamefont {Ross},\ and\ \citenamefont {Zegers}}]{Zegers}%
  \BibitemOpen
  \bibfield  {author} {\bibinfo {author} {\bibfnamefont {R.}~\bibnamefont
  {Gregory}}, \bibinfo {author} {\bibfnamefont {S.~F.}\ \bibnamefont {Ross}}, \
  and\ \bibinfo {author} {\bibfnamefont {R.}~\bibnamefont {Zegers}},\ }\href
  {\doibase 10.1088/1126-6708/2008/09/029} {\bibfield  {journal} {\bibinfo
  {journal} {JHEP}\ }\textbf {\bibinfo {volume} {09}},\ \bibinfo {pages} {029}
  (\bibinfo {year} {2008})},\ \Eprint {http://arxiv.org/abs/0802.2037}
  {arXiv:0802.2037 [hep-th]} \BibitemShut {NoStop}%
\bibitem [{\citenamefont {Heydari-Fard}\ and\ \citenamefont
  {Sepangi}(2009)}]{Heydari}%
  \BibitemOpen
  \bibfield  {author} {\bibinfo {author} {\bibfnamefont {M.}~\bibnamefont
  {Heydari-Fard}}\ and\ \bibinfo {author} {\bibfnamefont {H.~R.}\ \bibnamefont
  {Sepangi}},\ }\href {\doibase 10.1088/1475-7516/2009/02/029} {\bibfield
  {journal} {\bibinfo  {journal} {JCAP}\ }\textbf {\bibinfo {volume} {02}},\
  \bibinfo {pages} {029} (\bibinfo {year} {2009})},\ \Eprint
  {http://arxiv.org/abs/0903.0066} {arXiv:0903.0066 [gr-qc]} \BibitemShut
  {NoStop}%
\bibitem [{\citenamefont {Dai}\ and\ \citenamefont {Stojkovic}(2011)}]{Dai}%
  \BibitemOpen
  \bibfield  {author} {\bibinfo {author} {\bibfnamefont {D.-C.}\ \bibnamefont
  {Dai}}\ and\ \bibinfo {author} {\bibfnamefont {D.}~\bibnamefont
  {Stojkovic}},\ }\href {\doibase 10.1016/j.physletb.2011.09.038} {\bibfield
  {journal} {\bibinfo  {journal} {Phys. Lett. B}\ }\textbf {\bibinfo {volume}
  {704}},\ \bibinfo {pages} {354} (\bibinfo {year} {2011})},\ \Eprint
  {http://arxiv.org/abs/1004.3291} {arXiv:1004.3291 [gr-qc]} \BibitemShut
  {NoStop}%
\bibitem [{\citenamefont {Yoshino}(2009)}]{Yoshino}%
  \BibitemOpen
  \bibfield  {author} {\bibinfo {author} {\bibfnamefont {H.}~\bibnamefont
  {Yoshino}},\ }\href {\doibase 10.1088/1126-6708/2009/01/068} {\bibfield
  {journal} {\bibinfo  {journal} {JHEP}\ }\textbf {\bibinfo {volume} {01}},\
  \bibinfo {pages} {068} (\bibinfo {year} {2009})},\ \Eprint
  {http://arxiv.org/abs/0812.0465} {arXiv:0812.0465 [gr-qc]} \BibitemShut
  {NoStop}%
\bibitem [{\citenamefont {Kleihaus}\ \emph {et~al.}(2011)\citenamefont
  {Kleihaus}, \citenamefont {Kunz}, \citenamefont {Radu},\ and\ \citenamefont
  {Senkbeil}}]{Kleihaus}%
  \BibitemOpen
  \bibfield  {author} {\bibinfo {author} {\bibfnamefont {B.}~\bibnamefont
  {Kleihaus}}, \bibinfo {author} {\bibfnamefont {J.}~\bibnamefont {Kunz}},
  \bibinfo {author} {\bibfnamefont {E.}~\bibnamefont {Radu}}, \ and\ \bibinfo
  {author} {\bibfnamefont {D.}~\bibnamefont {Senkbeil}},\ }\href {\doibase
  10.1103/PhysRevD.83.104050} {\bibfield  {journal} {\bibinfo  {journal} {Phys.
  Rev. D}\ }\textbf {\bibinfo {volume} {83}},\ \bibinfo {pages} {104050}
  (\bibinfo {year} {2011})},\ \Eprint {http://arxiv.org/abs/1103.4758}
  {arXiv:1103.4758 [gr-qc]} \BibitemShut {NoStop}%
\bibitem [{\citenamefont {Kuerten}\ and\ \citenamefont
  {da~Rocha}(2016)}]{daRocha2}%
  \BibitemOpen
  \bibfield  {author} {\bibinfo {author} {\bibfnamefont {A.~M.}\ \bibnamefont
  {Kuerten}}\ and\ \bibinfo {author} {\bibfnamefont {R.~a.}\ \bibnamefont
  {da~Rocha}},\ }\href {\doibase 10.1007/s10714-016-2092-8} {\bibfield
  {journal} {\bibinfo  {journal} {Gen. Rel. Grav.}\ }\textbf {\bibinfo {volume}
  {48}},\ \bibinfo {pages} {90} (\bibinfo {year} {2016})},\ \Eprint
  {http://arxiv.org/abs/1407.1483} {arXiv:1407.1483 [gr-qc]} \BibitemShut
  {NoStop}%
\bibitem [{\citenamefont {Andrianov}\ and\ \citenamefont
  {Kurkov}(2010)}]{Andrianov1}%
  \BibitemOpen
  \bibfield  {author} {\bibinfo {author} {\bibfnamefont {A.~A.}\ \bibnamefont
  {Andrianov}}\ and\ \bibinfo {author} {\bibfnamefont {M.~A.}\ \bibnamefont
  {Kurkov}},\ }\href {\doibase 10.1103/PhysRevD.82.104027} {\bibfield
  {journal} {\bibinfo  {journal} {Phys. Rev. D}\ }\textbf {\bibinfo {volume}
  {82}},\ \bibinfo {pages} {104027} (\bibinfo {year} {2010})},\ \Eprint
  {http://arxiv.org/abs/1008.2705} {arXiv:1008.2705 [hep-th]} \BibitemShut
  {NoStop}%
\bibitem [{\citenamefont {Andrianov}\ and\ \citenamefont
  {Kurkov}(2011)}]{Andrianov2}%
  \BibitemOpen
  \bibfield  {author} {\bibinfo {author} {\bibfnamefont {A.~A.}\ \bibnamefont
  {Andrianov}}\ and\ \bibinfo {author} {\bibfnamefont {M.~A.}\ \bibnamefont
  {Kurkov}},\ }\href {\doibase 10.1007/s11232-011-0140-9} {\bibfield  {journal}
  {\bibinfo  {journal} {Theor. Math. Phys.}\ }\textbf {\bibinfo {volume}
  {169}},\ \bibinfo {pages} {1629} (\bibinfo {year} {2011})}\BibitemShut
  {NoStop}%
\bibitem [{\citenamefont {Cuadros-Melgar}\ \emph {et~al.}(2008)\citenamefont
  {Cuadros-Melgar}, \citenamefont {Papantonopoulos}, \citenamefont
  {Tsoukalas},\ and\ \citenamefont {Zamarias}}]{Cuadros}%
  \BibitemOpen
  \bibfield  {author} {\bibinfo {author} {\bibfnamefont {B.}~\bibnamefont
  {Cuadros-Melgar}}, \bibinfo {author} {\bibfnamefont {E.}~\bibnamefont
  {Papantonopoulos}}, \bibinfo {author} {\bibfnamefont {M.}~\bibnamefont
  {Tsoukalas}}, \ and\ \bibinfo {author} {\bibfnamefont {V.}~\bibnamefont
  {Zamarias}},\ }\href {\doibase 10.1103/PhysRevLett.100.221601} {\bibfield
  {journal} {\bibinfo  {journal} {Phys. Rev. Lett.}\ }\textbf {\bibinfo
  {volume} {100}},\ \bibinfo {pages} {221601} (\bibinfo {year} {2008})},\
  \Eprint {http://arxiv.org/abs/0712.3232} {arXiv:0712.3232 [hep-th]}
  \BibitemShut {NoStop}%
\bibitem [{\citenamefont {Kanti}\ \emph {et~al.}(2013)\citenamefont {Kanti},
  \citenamefont {Pappas},\ and\ \citenamefont {Zuleta}}]{KPZ}%
  \BibitemOpen
  \bibfield  {author} {\bibinfo {author} {\bibfnamefont {P.}~\bibnamefont
  {Kanti}}, \bibinfo {author} {\bibfnamefont {N.}~\bibnamefont {Pappas}}, \
  and\ \bibinfo {author} {\bibfnamefont {K.}~\bibnamefont {Zuleta}},\ }\href
  {\doibase 10.1088/0264-9381/30/23/235017} {\bibfield  {journal} {\bibinfo
  {journal} {Class. Quant. Grav.}\ }\textbf {\bibinfo {volume} {30}},\ \bibinfo
  {pages} {235017} (\bibinfo {year} {2013})},\ \Eprint
  {http://arxiv.org/abs/1309.7642} {arXiv:1309.7642 [hep-th]} \BibitemShut
  {NoStop}%
\bibitem [{\citenamefont {Kanti}\ \emph {et~al.}(2016)\citenamefont {Kanti},
  \citenamefont {Pappas},\ and\ \citenamefont {Pappas}}]{KPP}%
  \BibitemOpen
  \bibfield  {author} {\bibinfo {author} {\bibfnamefont {P.}~\bibnamefont
  {Kanti}}, \bibinfo {author} {\bibfnamefont {N.}~\bibnamefont {Pappas}}, \
  and\ \bibinfo {author} {\bibfnamefont {T.}~\bibnamefont {Pappas}},\ }\href
  {\doibase 10.1088/0264-9381/33/1/015003} {\bibfield  {journal} {\bibinfo
  {journal} {Class. Quant. Grav.}\ }\textbf {\bibinfo {volume} {33}},\ \bibinfo
  {pages} {015003} (\bibinfo {year} {2016})},\ \Eprint
  {http://arxiv.org/abs/1507.02625} {arXiv:1507.02625 [hep-th]} \BibitemShut
  {NoStop}%
\bibitem [{\citenamefont {Kanti}\ \emph {et~al.}(2018)\citenamefont {Kanti},
  \citenamefont {Nakas},\ and\ \citenamefont {Pappas}}]{KNP1}%
  \BibitemOpen
  \bibfield  {author} {\bibinfo {author} {\bibfnamefont {P.}~\bibnamefont
  {Kanti}}, \bibinfo {author} {\bibfnamefont {T.}~\bibnamefont {Nakas}}, \ and\
  \bibinfo {author} {\bibfnamefont {N.}~\bibnamefont {Pappas}},\ }\href
  {\doibase 10.1103/PhysRevD.98.064025} {\bibfield  {journal} {\bibinfo
  {journal} {Phys. Rev. D}\ }\textbf {\bibinfo {volume} {98}},\ \bibinfo
  {pages} {064025} (\bibinfo {year} {2018})},\ \Eprint
  {http://arxiv.org/abs/1807.06880} {arXiv:1807.06880 [gr-qc]} \BibitemShut
  {NoStop}%
\bibitem [{\citenamefont {Nakas}\ \emph {et~al.}(2019)\citenamefont {Nakas},
  \citenamefont {Pappas},\ and\ \citenamefont {Kanti}}]{KNP2}%
  \BibitemOpen
  \bibfield  {author} {\bibinfo {author} {\bibfnamefont {T.}~\bibnamefont
  {Nakas}}, \bibinfo {author} {\bibfnamefont {N.}~\bibnamefont {Pappas}}, \
  and\ \bibinfo {author} {\bibfnamefont {P.}~\bibnamefont {Kanti}},\ }\href
  {\doibase 10.1103/PhysRevD.99.124040} {\bibfield  {journal} {\bibinfo
  {journal} {Phys. Rev. D}\ }\textbf {\bibinfo {volume} {99}},\ \bibinfo
  {pages} {124040} (\bibinfo {year} {2019})},\ \Eprint
  {http://arxiv.org/abs/1904.00216} {arXiv:1904.00216 [hep-th]} \BibitemShut
  {NoStop}%
\bibitem [{\citenamefont {Nakas}\ \emph {et~al.}(2020)\citenamefont {Nakas},
  \citenamefont {Kanti},\ and\ \citenamefont {Pappas}}]{KNP3}%
  \BibitemOpen
  \bibfield  {author} {\bibinfo {author} {\bibfnamefont {T.}~\bibnamefont
  {Nakas}}, \bibinfo {author} {\bibfnamefont {P.}~\bibnamefont {Kanti}}, \ and\
  \bibinfo {author} {\bibfnamefont {N.}~\bibnamefont {Pappas}},\ }\href
  {\doibase 10.1103/PhysRevD.101.084056} {\bibfield  {journal} {\bibinfo
  {journal} {Phys. Rev. D}\ }\textbf {\bibinfo {volume} {101}},\ \bibinfo
  {pages} {084056} (\bibinfo {year} {2020})},\ \Eprint
  {http://arxiv.org/abs/2001.07226} {arXiv:2001.07226 [hep-th]} \BibitemShut
  {NoStop}%
\bibitem [{\citenamefont {Morris}\ and\ \citenamefont {Thorne}(1988)}]{MT}%
  \BibitemOpen
  \bibfield  {author} {\bibinfo {author} {\bibfnamefont {M.}~\bibnamefont
  {Morris}}\ and\ \bibinfo {author} {\bibfnamefont {K.}~\bibnamefont
  {Thorne}},\ }\href {\doibase 10.1119/1.15620} {\bibfield  {journal} {\bibinfo
   {journal} {Am. J. Phys.}\ }\textbf {\bibinfo {volume} {56}},\ \bibinfo
  {pages} {395} (\bibinfo {year} {1988})}\BibitemShut {NoStop}%
\bibitem [{\citenamefont {Israel}(1966)}]{Israel}%
  \BibitemOpen
  \bibfield  {author} {\bibinfo {author} {\bibfnamefont {W.}~\bibnamefont
  {Israel}},\ }\href {\doibase 10.1007/BF02710419} {\bibfield  {journal}
  {\bibinfo  {journal} {Nuovo Cim. B}\ }\textbf {\bibinfo {volume} {44S10}},\
  \bibinfo {pages} {1} (\bibinfo {year} {1966})},\ \bibinfo {note} {[Erratum:
  Nuovo Cim.B 48, 463 (1967)]}\BibitemShut {NoStop}%
\bibitem [{\citenamefont {Shiromizu}\ \emph {et~al.}(2000)\citenamefont
  {Shiromizu}, \citenamefont {Maeda},\ and\ \citenamefont {Sasaki}}]{SMS}%
  \BibitemOpen
  \bibfield  {author} {\bibinfo {author} {\bibfnamefont {T.}~\bibnamefont
  {Shiromizu}}, \bibinfo {author} {\bibfnamefont {K.-i.}\ \bibnamefont
  {Maeda}}, \ and\ \bibinfo {author} {\bibfnamefont {M.}~\bibnamefont
  {Sasaki}},\ }\href {\doibase 10.1103/PhysRevD.62.024012} {\bibfield
  {journal} {\bibinfo  {journal} {Phys. Rev. D}\ }\textbf {\bibinfo {volume}
  {62}},\ \bibinfo {pages} {024012} (\bibinfo {year} {2000})},\ \Eprint
  {http://arxiv.org/abs/gr-qc/9910076} {arXiv:gr-qc/9910076} \BibitemShut
  {NoStop}%
\bibitem [{\citenamefont {Maldacena}(1999)}]{Maldacena}%
  \BibitemOpen
  \bibfield  {author} {\bibinfo {author} {\bibfnamefont {J.~M.}\ \bibnamefont
  {Maldacena}},\ }\href {\doibase 10.1023/A:1026654312961} {\bibfield
  {journal} {\bibinfo  {journal} {Int. J. Theor. Phys.}\ }\textbf {\bibinfo
  {volume} {38}},\ \bibinfo {pages} {1113} (\bibinfo {year} {1999})},\ \Eprint
  {http://arxiv.org/abs/hep-th/9711200} {arXiv:hep-th/9711200} \BibitemShut
  {NoStop}%
\bibitem [{\citenamefont {Gubser}\ \emph {et~al.}(1998)\citenamefont {Gubser},
  \citenamefont {Klebanov},\ and\ \citenamefont {Polyakov}}]{Gubser}%
  \BibitemOpen
  \bibfield  {author} {\bibinfo {author} {\bibfnamefont {S.}~\bibnamefont
  {Gubser}}, \bibinfo {author} {\bibfnamefont {I.~R.}\ \bibnamefont
  {Klebanov}}, \ and\ \bibinfo {author} {\bibfnamefont {A.~M.}\ \bibnamefont
  {Polyakov}},\ }\href {\doibase 10.1016/S0370-2693(98)00377-3} {\bibfield
  {journal} {\bibinfo  {journal} {Phys. Lett. B}\ }\textbf {\bibinfo {volume}
  {428}},\ \bibinfo {pages} {105} (\bibinfo {year} {1998})},\ \Eprint
  {http://arxiv.org/abs/hep-th/9802109} {arXiv:hep-th/9802109} \BibitemShut
  {NoStop}%
\bibitem [{\citenamefont {Witten}(1998)}]{Witten}%
  \BibitemOpen
  \bibfield  {author} {\bibinfo {author} {\bibfnamefont {E.}~\bibnamefont
  {Witten}},\ }\href {\doibase 10.4310/ATMP.1998.v2.n2.a2} {\bibfield
  {journal} {\bibinfo  {journal} {Adv. Theor. Math. Phys.}\ }\textbf {\bibinfo
  {volume} {2}},\ \bibinfo {pages} {253} (\bibinfo {year} {1998})},\ \Eprint
  {http://arxiv.org/abs/hep-th/9802150} {arXiv:hep-th/9802150} \BibitemShut
  {NoStop}%
\bibitem [{\citenamefont {Da~Rold}\ and\ \citenamefont
  {Pomarol}(2005)}]{Pomarol}%
  \BibitemOpen
  \bibfield  {author} {\bibinfo {author} {\bibfnamefont {L.}~\bibnamefont
  {Da~Rold}}\ and\ \bibinfo {author} {\bibfnamefont {A.}~\bibnamefont
  {Pomarol}},\ }\href {\doibase 10.1016/j.nuclphysb.2005.05.009} {\bibfield
  {journal} {\bibinfo  {journal} {Nucl. Phys. B}\ }\textbf {\bibinfo {volume}
  {721}},\ \bibinfo {pages} {79} (\bibinfo {year} {2005})},\ \Eprint
  {http://arxiv.org/abs/hep-ph/0501218} {arXiv:hep-ph/0501218} \BibitemShut
  {NoStop}%
\bibitem [{\citenamefont {Alho}\ \emph {et~al.}(2013)\citenamefont {Alho},
  \citenamefont {Evans},\ and\ \citenamefont {Tuominen}}]{Alho}%
  \BibitemOpen
  \bibfield  {author} {\bibinfo {author} {\bibfnamefont {T.}~\bibnamefont
  {Alho}}, \bibinfo {author} {\bibfnamefont {N.}~\bibnamefont {Evans}}, \ and\
  \bibinfo {author} {\bibfnamefont {K.}~\bibnamefont {Tuominen}},\ }\href
  {\doibase 10.1103/PhysRevD.88.105016} {\bibfield  {journal} {\bibinfo
  {journal} {Phys. Rev. D}\ }\textbf {\bibinfo {volume} {88}},\ \bibinfo
  {pages} {105016} (\bibinfo {year} {2013})},\ \Eprint
  {http://arxiv.org/abs/1307.4896} {arXiv:1307.4896 [hep-ph]} \BibitemShut
  {NoStop}%
\bibitem [{\citenamefont {Ballon~Bayona}\ \emph {et~al.}(2008)\citenamefont
  {Ballon~Bayona}, \citenamefont {Boschi-Filho}, \citenamefont {Braga},\ and\
  \citenamefont {Pando~Zayas}}]{Ballon}%
  \BibitemOpen
  \bibfield  {author} {\bibinfo {author} {\bibfnamefont {C.}~\bibnamefont
  {Ballon~Bayona}}, \bibinfo {author} {\bibfnamefont {H.}~\bibnamefont
  {Boschi-Filho}}, \bibinfo {author} {\bibfnamefont {N.~R.}\ \bibnamefont
  {Braga}}, \ and\ \bibinfo {author} {\bibfnamefont {L.~A.}\ \bibnamefont
  {Pando~Zayas}},\ }\href {\doibase 10.1103/PhysRevD.77.046002} {\bibfield
  {journal} {\bibinfo  {journal} {Phys. Rev. D}\ }\textbf {\bibinfo {volume}
  {77}},\ \bibinfo {pages} {046002} (\bibinfo {year} {2008})},\ \Eprint
  {http://arxiv.org/abs/0705.1529} {arXiv:0705.1529 [hep-th]} \BibitemShut
  {NoStop}%
\end{thebibliography}%
\bibliographystyle{apsrev4-1}

\end{document}